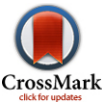

# Efficient Transfer Entropy Analysis of Non-Stationary Neural Time Series

**Patricia Wollstadt**[1]*[,9], **Mario Martínez-Zarzuela**[4,9], **Raul Vicente**[2,3], **Francisco J. Díaz-Pernas**[4], **Michael Wibral**[1]

1 MEG Unit, Brain Imaging Center, Goethe University, Frankfurt, Germany, 2 Frankfurt Institute for Advanced Studies (FIAS), Goethe University, Frankfurt, Germany, 3 Max-Planck Institute for Brain Research, Frankfurt, Germany, 4 Department of Signal Theory and Communications and Telematics Engineering, University of Valladolid, Valladolid, Spain

## Abstract

Information theory allows us to investigate information processing in neural systems in terms of information transfer, storage and modification. Especially the measure of information transfer, transfer entropy, has seen a dramatic surge of interest in neuroscience. Estimating transfer entropy from two processes requires the observation of multiple realizations of these processes to estimate associated probability density functions. To obtain these necessary observations, available estimators typically assume stationarity of processes to allow pooling of observations over time. This assumption however, is a major obstacle to the application of these estimators in neuroscience as observed processes are often non-stationary. As a solution, Gomez-Herrero and colleagues theoretically showed that the stationarity assumption may be avoided by estimating transfer entropy from an ensemble of realizations. Such an ensemble of realizations is often readily available in neuroscience experiments in the form of experimental trials. Thus, in this work we combine the ensemble method with a recently proposed transfer entropy estimator to make transfer entropy estimation applicable to non-stationary time series. We present an efficient implementation of the approach that is suitable for the increased computational demand of the ensemble method's practical application. In particular, we use a massively parallel implementation for a graphics processing unit to handle the computationally most heavy aspects of the ensemble method for transfer entropy estimation. We test the performance and robustness of our implementation on data from numerical simulations of stochastic processes. We also demonstrate the applicability of the ensemble method to magnetoencephalographic data. While we mainly evaluate the proposed method for neuroscience data, we expect it to be applicable in a variety of fields that are concerned with the analysis of information transfer in complex biological, social, and artificial systems.

**Citation:** Wollstadt P, Martínez-Zarzuela M, Vicente R, Díaz-Pernas FJ, Wibral M (2014) Efficient Transfer Entropy Analysis of Non-Stationary Neural Time Series. PLoS ONE 9(7): e102833. doi:10.1371/journal.pone.0102833

**Editor:** Daniele Marinazzo, Universiteit Gent, Belgium

**Received** December 20, 2013; **Accepted** June 24, 2014; **Published** July 28, 2014

**Copyright:** © 2014 Wollstadt et al. This is an open-access article distributed under the terms of the Creative Commons Attribution License, which permits unrestricted use, distribution, and reproduction in any medium, provided the original author and source are credited.

**Funding:** MW and RV received financial support from LOEWE Grant "Neuronale Koordination Forschungsschwerpunkt Frankfurt (NeFF)". MMZ received financial support from the University of Valladolid. The funders had no role in study design, data collection and analysis, decision to publish, or preparation of the manuscript.

**Competing Interests:** The authors have declared that no competing interests exist.

* Email: p.wollstadt@stud.uni-frankfurt.de

9 These authors contributed equally to this work.

## Introduction

We typically think of the brain as some kind of information processing system, albeit mostly without having a strict definition of information processing in mind. However, more formal accounts of information processing exist, and may be applied to brain research. In efforts dating back to Alan Turing [1] it was shown that any act of information processing can be broken down into the three components of information storage, information transfer, and information modification [1–4]. These components can be easily identified in theoretical or technical information processing systems, such as ordinary computers, based on the specialized machinery for and the spatial separation of these component functions. In these examples, a separation of the components of information processing via a specialized mathematical formalism seems almost superfluous. However, in biological systems in general, and in the brain in particular, we deal with a form of distributed information processing based on a large number of interacting agents (neurons), and each agent at each moment in time subserves any of the three component functions to a varying degree (see [5] for an example of time-varying storage). In neural systems it is indeed crucial to understand where and when information storage, transfer and modification take place, to constrain possible algorithms run by the system. While there is still a struggle to properly define information modification [6,7] and its proper measure [8–12], well established measures for (local active) information storage [13], information transfer [14], and its localization in time and space [15,16] exist, and are applied in neuroscience (for information storage see [5,17,18], for information transfer see below).

Especially the measure for information transfer, transfer entropy (TE), has seen a dramatic surge of interest in neuroscience [19–41], physiology [42–44], and other fields [6,15,31,45,46]. Nevertheless, conceptual and practical problems still exist. On the conceptual side, information transfer has been for a while confused with causal interactions, and only some recent studies [47–49] made clear that there can be no one-to-one mapping between causal interactions and information transfer, because causal interactions will subserve all *three* components of information processing (transfer, storage, modification). However, it is information transfer, rather than causal interactions, we might be interested in when trying to understand a computational process in the brain [48].

On the practical side, efforts to apply measures of information transfer in neuroscience have been hampered by two obstacles: (1) the need to analyze the information processing in a multivariate manner, to arrive at unambiguous conclusions that are not clouded by spurious traces of information transfer, e.g. due to effects of cascades and common drivers; (2) the fact that available estimators of information transfer typically require the processes under investigation to be stationary.

The first obstacle can in principle be overcome by conditioning TE on all other processes in a system, using a fully multivariate approach that had already been formulated by Schreiber [14]. However, the naive application of this approach normally fails because the samples available for estimation are typically too few. Therefore, recently four approaches to build an approximate representation of the information transfer network have been suggested: Lizier and Rubinov [50], Faes and colleagues [44], and Stramaglia and colleagues [51] presented algorithms for iterative inclusion of processes into an approximate multivariate description. In the approach suggested by Stramaglia and colleagues, conditional mutual information terms are additionally computed at each level as a self-truncating series expansion, following a suggestion by Bettencourt and colleagues [52]. In contrast to these approaches that explicitly compute conditional TE terms, we recently suggested an approximation based on a reconstruction of information transfer delays [53] and a graphical pruning algorithm [54]. While the first three approaches will eventually be closer to the ground truth, the graphical method may be better applicable to very limited amounts of data. In sum, the first problem of multivariate analysis can be considered solved for practical purposes, given enough data are available.

The second obstacle of dealing with non-stationary processes is also not a fundamental one, as the definition of TE relies on the availability of multiple realizations of (two or more) random processes, that can be obtained by running an ensemble of many identical copies of the processes in question, or by running one process multiple times. Only when obtaining data from such copies or repetitions is impossible, we have to turn to a stationarity assumption in order to evaluate the necessary probability density functions (PDF) based on a single realization.

Fortunately, in neuroscience we can often obtain many realizations of the processes in question by repeating an experiment. In fact, this is the typical procedure in neuroscience - we repeat trials under conditions that are kept as constant as possible (i.e we create a cyclostationary process). The possibility to use such an *ensemble* of data to estimate the time resolved TE has already been demonstrated theoretically by Gomez-Herrero and colleagues [55]. Practically, however, the statistical testing necessary for this ensemble-based method leads to an increase in computational cost by several orders of magnitude, as some shortcuts in statistical validation that can be taken for stationary data cannot be used for the ensemble approach (see [56]): For stationary data, TE is calculated per trial and *one* set of trial-based surrogate data may be used for statistical testing. The ensemble method does not allow for trial-based TE estimation as TE is estimated across trials. Instead, the ensemble method requires the generation of a sufficiently large number of surrogate data sets, for *all* of which TE has to be estimated, thus multiplying the computational demand by the number of surrogate data sets.

Therefore, the use of the ensemble method has remained a theoretical possibility so far, especially in combination with the nearest neighbor-based estimation techniques by Kraskov and colleagues [57] that provide the most precise, yet computationally most heavy TE estimates. For example, the analysis of magnetoencephalographic data presented here would require a runtime of 8200 h for 15 subjects and a single experimental condition. It is easy to see that any practical application of the methods hinges on a substantial speed-up of the computation.

Fortunately, the algorithms involved in ensemble-based TE estimation, lend themselves easily to data-parallel processing, since most of the algorithm's fundamental parts can be computed simultaneously. Thus, our problem matches the massively parallel architecture of Graphics Processing Unit (GPU) devices well. GPUs were originally devised only for computer graphics, but are routinely used to speed up computations in many areas today [58,59]. Also in neuroscience, where applied algorithms continue to grow faster in complexity than the CPU performance, the use of GPUs with data-parallel methods is becoming increasingly important [60] and GPUs have successfully been used to speedup time series analysis in neuroscientific experiments [61–66].

Thus, in order to overcome the limitations set by the computational demands of TE analysis from an ensemble of data, we developed a GPU implementation of the algorithm, where the neighbor searches underlying the binless TE estimation [57] are executed in parallel on the GPU. After parallelizing this computationally most heavy aspect of TE estimation we were able to use the ensemble method for TE estimation proposed by [55], to estimate time-resolved TE from non-stationary neural time-series in acceptable time. Using the new GPU-based TE estimation tool on a high-end consumer graphics card reduced computation time by a factor of 50 compared to the CPU optimized TE search used previously [67]. In practical terms, this speedup shortens the duration of an ensemble-based analysis for typical neural data sets enough to make the application of the ensemble method feasible for the first time.

## Background

Our study focuses on making the application of ensemble-based estimation of TE from non-stationary data practical using a GPU-based algorithm. For the convenience of the reader, we will also present the necessary background on stationarity, TE estimation using the Kraskov-Stögbauer-Grassberger (KSG) estimator [19], and the ensemble method of Gomez-Herrero et al. [55] in condensed form in a short background section below. Readers well familiar with these topics can safely skip ahead to the *Implementation* section below.

## Notation

To describe practical TE estimation from time series recorded in a system of interest $\mathcal{X}$ (e.g. a brain area), we first have to formalize these recordings mathematically: We define an observed time series $\mathbf{x} = (x_1, x_2, \ldots, x_t, \ldots, x_N)$ as a realization of a random process $\mathbf{X} = (X_1, X_2, \ldots, X_t, \ldots, X_N)$. A random process here is simply a collection of individual random variables sorted by an integer index $t \in \{1, \ldots, N\}$, representing time. TE or other information theoretic functionals are then calculated from the random variables' joint PDFs $p_{X_s Y_t}(X_s = a_i, Y_t = b_j)$ and conditional PDFs $p_{X_s | Y_t}(X_s = a_i | Y_t = b_j)$ (with $s, t \in \{1, \ldots, N\}$), where $\mathcal{A}_{X_s} = \{a_1, a_2, \ldots, a_i, \ldots, a_I\}$ and $\mathcal{B}_{Y_t} = \{b_1, b_2, \ldots, b_j, \ldots, b_J\}$ are all possible outcomes of the random variables $X_s$ and $Y_t$, and where $p_{X_s | Y_t}(X_s = a_i | Y_t = b_j) = \dfrac{p_{X_s Y_t}(X_s = a_i, Y_t = b_j)}{p_{Y_t}(Y_t = b_j)}$.

We call information theoretic quantities functionals as they are defined as functions that map from the space of PDFs to the real numbers. If we have to estimate the underlying probabilities from experimental data first, the mapping from the data to the information theoretic quantity (a real number) is called an estimator.

## Stationarity and non-stationarity in experimental time series

PDFs in neuroscience are typically not known *a priori*, so in order to estimate information theoretic functionals, these PDFs have to be reconstructed from a sufficient amount of observed realizations of the process. How these realizations are obtained from data depends on whether the process in question is stationary or non-stationary. Stationarity of a process means that PDFs of the random variables that form the random process do not change over time, such that $p_{X_t}(X_t = a_j) = p_{X_{t'}}(X_{t'} = a_j), \forall t, t' \in \mathbb{N}$. Any PDF $p_{X_t}(\cdot)$ may be estimated from one observation of process X by means of collecting realizations $x_{t'}$ *over time* $t' \in \{1, \ldots, N\}$.

For processes that do not fulfill the stationarity-assumption, temporal pooling is not applicable as PDFs vary over time $t$ and some random variables $X_t$, $X_s$ (at least two) are associated with different PDFs $p_{X_t}(\cdot)$, $p_{X_s}(\cdot)$ (Figure 1). To still gain the necessary multiple observations of a random variable $X_t$ we may resort to either run multiple physical copies of the process X or – in cases where physical copies are unavailable – we may repeat a process in time. If we choose the repetition large enough, i.e. there is a sufficiently large set $\mathcal{R}$ of time points $\Theta$, at which the process is repeated, we can assume that

$$\exists \mathcal{R} \subseteq \mathbb{N} \land \mathcal{R} \neq \emptyset : p_{X_{\Theta+t}}(a_j) = p_{X_{\Theta'+t}}(a_j)$$
$$\forall t \in \mathbb{N} : t < min(|\Theta - \Theta'|), \forall \Theta, \Theta' \in \mathcal{R}, \forall a_j \in \mathcal{A}_{X_t}, \tag{1}$$

i.e. PDFs $p_{X_{\Theta+t}}(\cdot)$ at time point $t$ relative to the onset of the repetition at $\Theta$ are equal over all $R = |\mathcal{R}|$ repetitions. We call the repeated observations of a process an *ensemble* of time series. We may obtain a reliable estimation of $p_{X_{\Theta+t}}(\cdot)$ from this ensemble by evaluating $p_{\cdot}(\cdot)$ over all observations $x_{\Theta+t}, \forall \Theta \in \mathcal{R}$. For the sake of readability, we will refer to these observations from the ensemble as $x_t(r)$, where $t$ refers to a time point $t$, relative to the beginning of the process at time $\Theta$, and $r = 1, \ldots, R$ refers to the index of the repetition. If a process is repeated periodically, i.e. the repetitions are spaced by a fixed interval $T$, we call such a process cyclostationary [68]:

$$\exists T : \forall t \ p_{X_t}(a_j) = p_{X_{nT+t}}(a_j) \quad \forall n, t \in \mathbb{N}, t < T, \forall a_j \in \mathcal{A}_{X_t}. \tag{2}$$

In neuroscience, ensemble evaluation for the estimation of information theoretic functionals becomes relevant as physical copies of a process are typically not available and stationarity of a process can not necessarily be assumed. Gomez-Herrero and colleagues recently showed how ensemble averaging may be used to nevertheless estimate information theoretic functionals from cyclostationary processes [55]. In neuroscience for example, a cyclostationary process, and thus an ensemble of data, is obtained by repeating an experimental manipulation, e.g. the presentation of a stimulus; these repetitions are often called experimental *trials*. In the remainder of this article, we will use the term repetition, and interpret trials from a neuroscience experiment as a special case of repetitions of a random process. Building on such repetitions, we next demonstrate a computationally efficient approach to the estimation of TE using the ensemble method proposed in [55].

## Transfer entropy estimation from an ensemble of time series

**Ensemble-based TE functional.** When independent repetitions of an experimental condition are available, it is possible to use ensemble evaluation to estimate various PDFs from an ensemble of repetitions of the time series [55]. By eliminating the need for pooling data over time, and instead pooling over repetitions, ensemble methods can be used to estimate information theoretic functionals for non-stationary time series. Here, we follow the approach of [55] and present an ensemble TE functional that extends the TE functional presented in [19,20,53] and also takes into account an extension of the original formulation of TE, presented in [53], guaranteeing self prediction optimality (indicated by the subscript *SPO*). In the next subsection, we will then present a practical and data-efficient estimator of this functional. The functional reads

$$TE_{SPO}(X \to Y, t, u) = I(Y_t; \mathbf{X}_{t-u}^{d_X} | \mathbf{Y}_{t-1}^{d_Y}), \tag{3}$$

where $I(\cdot; \cdot | \cdot)$ is the conditional mutual information, and $Y_t$, $\mathbf{Y}_{t-1}^{d_Y}$, and $\mathbf{X}_{t-u}^{d_X}$ are the current value and the $d_Y$-dimensional past state variables of the target process Y, and the $d_X$-dimensional past state variable at time $t-u$ of the source process X, respectively (see next paragraph for an explanation of states).

Rewriting this, taking into account repetitions $r$ of the random processes explicitly we obtain:

$$TE_{SPO}(X \to Y, t, u) = \sum_{\substack{y_t(r), \mathbf{y}_{t-1}^{d_Y}(r), \mathbf{x}_{t-u}^{d_X}(r) \\ \in \mathcal{A}_{Y_t, \mathbf{Y}_{t-1}^{d_Y}, \mathbf{X}_{t-u}^{d_X}}}} p\left(y_t(r), \mathbf{y}_{t-1}^{d_Y}(r), \mathbf{x}_{t-u}^{d_X}(r)\right)$$
$$\log \frac{p\left(y_t(r) | \mathbf{y}_{t-1}^{d_Y}(r), \mathbf{x}_{t-u}^{d_X}(r)\right)}{p\left(y_t(r) | \mathbf{y}_{t-1}^{d_Y}(r)\right)} \tag{4}$$

Here, $u$ is the assumed delay of the information transfer between processes $X$ and $Y$ [53]; $y_t(r)$ denotes the future observation of $Y$ in repetition $r = 1, \ldots, R$; $\mathbf{y}_{t-1}^{d_Y}(r)$ denotes the past state of $Y$ in repetition $r$ and $\mathbf{x}_{t-u}^{d_X}(r)$ denotes the past state of $X$ in repetition $r$. Note, that the functional $TE_{SPO}$ used here is a modified form of the original TE formulation introduced by Schreiber [14]. Schreiber defined TE as a conditional mutual information $TE(X \to Y, t) = I(Y_t; \mathbf{X}_{t-1}^{d_x} | \mathbf{Y}_{t-1}^{d_y})$, whereas the functional in eq. 3 implements the conditional mutual information $TE_{SPO}(X \to Y, t, u) = I(Y_t; \mathbf{X}_{t-u}^{d_x} | \mathbf{Y}_{t-1}^{d_y})$ [53]. The latter functional, $TE_{SPO}$, contains the definition of Schreiber as a special case for $u = 1$. Note that the two functionals are identical if $TE_{SPO}$ is used with the physically correct delay $\delta$ (i.e. $u = \delta$) and a proper embedding for the source, and the Schreiber measures is used with an over-embedding such that the source state at $(t - \delta)$ is still fully covered by the source embedding.

In addition to the original formulation of $TE_{SPO}$ in [53], here we explicitly state that the necessary realizations of the random variables in question are obtained through *ensemble evaluation* over repetitions $r$ – assuming the underlying processes to be repeatable or cyclostationary. Furthermore, we note explicitly that this ensemble-based functional introduces the possibility of time resolved TE estimates.

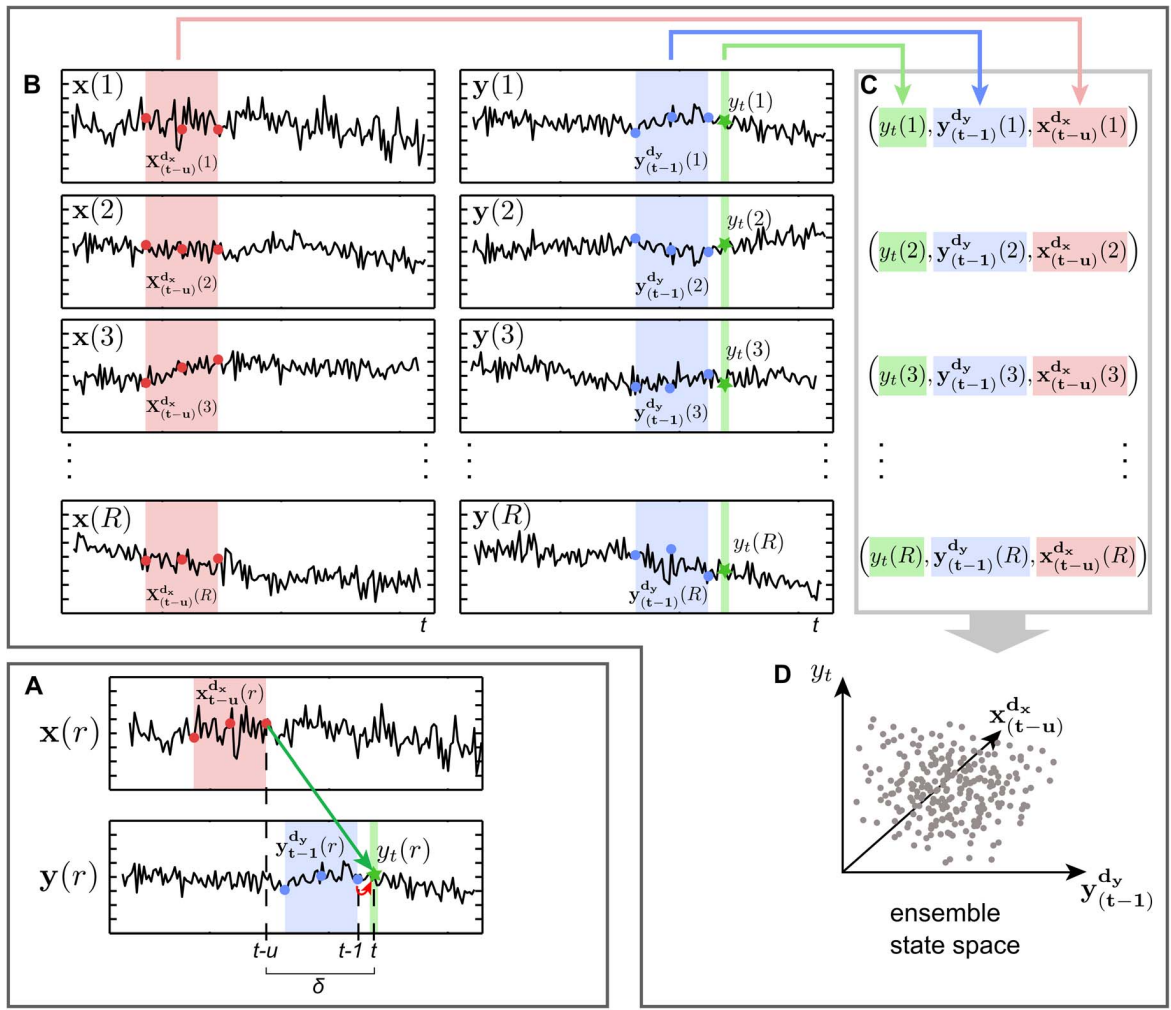

**Figure 1. Pooling of data over an ensemble of time series for transfer entropy (TE) estimation.** (A) Schematic account of TE. Two scalar time series $X_r$ and $Y_r$ recorded from the $r^{th}$ repetition of processes $X$ and $Y$, coupled with a delay $\delta$ (indicated by green arrow). Colored boxes indicate delay embedded states $\mathbf{x}_{t-u}^{d_X}(r)$, $\mathbf{y}_{t-1}^{d_Y}(r)$ for both time series with dimension $d_X = d_Y = 3$ samples (colored dots). The star on the $Y$ time series indicates the scalar observation $y_t$ that is obtained at the target time of information transfer $t$. The red arrow indicates self-information-transfer from the past of the target process to the random variable $Y_t$ at the target time. $u$ is chosen such that $u = \delta$ and influences of the state $\mathbf{x}_{t-u}^{d_X}(r)$ arrive exactly at the information target variable $Y_t$. Information in the past state of $X$ is useful to predict the future value of $Y$ and we obtain nonzero TE. (B) To estimate probability density functions for $\mathbf{x}_{t-u}^{d_X}(r)$, $\mathbf{y}_{t-1}^{d_Y}(r)$ and $y_t(r)$ at a certain point in time $t$, we collect their realizations from observed repetitions $r = 1, \ldots, R$. (C) Realizations for a single repetition are concatenated into one embedding vector and (D) combined into one ensemble state space. Note, that data are pooled over the ensemble of data instead of time. Nearest neighbor counts within the ensemble state space can then be used to derive TE using the Kraskov-estimator proposed in [57].
doi:10.1371/journal.pone.0102833.g001

We recently showed that the estimator presented in [53] can also be used to recover an unknown information transfer delay $\delta$ between two processes $X$ and $Y$, as $TE_{SPO}(X \rightarrow Y, t, u)$ is maximal when the assumed delay $u$ is equal to the true information transfer delay $\delta$ [53]. This holds for the extended estimator presented here, thus

$$\delta = \arg\max_u (TE_{SPO}(X \rightarrow Y, t, u)). \quad (5)$$

**State space reconstruction and practical estimator.** Transfer entropy differs from the lagged mutual information $I(Y_t; \mathbf{X}_{t-u}^{d_X})$ by the additional conditioning on the past of the target time series, $\mathbf{Y}_{t-1}^{d_Y}$. This additional conditioning serves two important functions. First, as mentioned already by Schreiber in the original paper [14], and later detailed by Lizier [4] and Wibral and colleagues [39,53], it removes the information about the future of the target time-series $Y_t$ that is already contained in its own past, $\mathbf{Y}_{t-1}^{d_Y}$. Second, this additional conditioning allows for a discovery of information transfer from the source $\mathbf{X}_{t-u}^{d_X}$ to the target that can only be seen when taking into account information from the past of the target $\mathbf{Y}_{t-1}^{d_Y}$ [69]. In the second case, the past information from the target serves to 'decode' this information transfer, and acts like a key in cryptography. As a consequence of this importance of the past of the target process it is very important to take all the necessary information in this past into account when evaluating the TE as in equation 4.

To this end we need to form a collection of past random variables

$$\mathbf{Y}_{t-1}^{d_Y} = (Y_{t-1}, Y_{t-1-\tau}, \ldots, Y_{t-1-(d_Y-1)\tau}), \tag{6}$$

such that their realizations,

$$\mathbf{y}_{t-1}^{d_Y}(r) = (y_{t-1}(r), y_{t-1-\tau}, \ldots, y_{t-1-(d_Y-1)\tau}), \tag{7}$$

are maximally informative about the future of the target process, $Y_t$.

This task is complicated by the fact the we often deal with multidimensional systems, of which we only observe a scalar variable (here modeled as our random processes X,Y). To see this, think for example of a pendulum (which is a two dimensional system) of which we record only the current position $Y_t$. If the pendulum is at its lowest point, it could be standing still, going left, or going right. To properly describe which state the pendulum is in, we need to know at least the realization of one more random variable $Y_{t-1}$ back in time. Collections of such past random variables whose realizations uniquely describe the state of a process are called *state variables*.

Such a sufficient collection of past variables, called a delay embedding vector, can always be reconstructed from scalar observations for low dimensional deterministic systems, such as the above pendulum, as shown by Takens [70]. Unfortunately, most real world systems are high-dimensional stochastic dynamic systems (best described by non-linear Langevin equations) rather than low-dimensional deterministic ones. For these systems it is not obvious that a delay embedding similar to Takens' approach would yield the desired results. In fact, many systems can be shown to require an infinite number of past random variables when only a scalar observable of the high-dimensional stochastic process is accessible. Nevertheless, as shown by Ragwitz and Kantz [71], the behavior of scalar observables of most of these systems can be approximated very well by a finite collection of such past variables for all practical purposes; in other words, these systems can be approximated well by a finite order, one-dimensional Markov-process.

For practical TE estimation using equation 4, we therefore proceed by first reconstructing the state variables of such approximated Markov processes for the two systems $\mathcal{X}$, $\mathcal{Y}$ from their scalar time series. Then, we use the statistics of nearest ensemble neighbors with a modified KSG estimator for TE evaluation [57].

Thus, we select a delay embedding vector of the form $\mathbf{Y}_{t-1}^{d_Y} = (Y_{t-1}, Y_{t-1-\tau}, \ldots, Y_{t-1-(d_Y-1)\tau})$ from equation 6 as our collection of past random variables – with realizations in repetition $r$ given by $\mathbf{y}_{t-1}^{d_Y}(r) = (y_{t-1}(r), y_{t-1-\tau}, \ldots, y_{t-1-(d_Y-1)\tau})$. Here, $d_Y$ is called the embedding dimension and $\tau$ the embedding delay. These embedding parameters $d_Y$ and $\tau$, are chosen such that they optimize a local predictor [71], as this avoids an overestimation of TE [53]; other approaches related to minimizing non-linear prediction errors are also possible [44]. In particular, $d_Y\cdot\tau$ is chosen such that $\mathbf{Y}_t^{d_Y}$ is conditionally independent of any $\mathbf{Y}_e^{d_Y}$ with $e < t - d\cdot\tau$ given $\mathbf{Y}_{t-1}^{d_Y}$. The same is done for the process X at time $t-u$.

Next, we decompose $TE_{SPO}$ into a sum of four individual Shannon entropies:

$$TE_{SPO}(X \to Y, t, u) = H\left(\mathbf{Y}_{t-1}^{d_Y}, \mathbf{X}_{t-u}^{d_X}\right) - H\left(Y_t, \mathbf{Y}_{t-1}^{d_Y}, \mathbf{X}_{t-u}^{d_X}\right) \\ + H\left(Y_t, \mathbf{Y}_{t-1}^{d_Y}\right) - H\left(\mathbf{Y}_{t-1}^{d_Y}\right), \tag{8}$$

The Shannon differential entropies in equation 8 can be estimated in a data efficient way using nearest neighbor techniques [72,73]. Nearest neighbor estimators yield a non-parametric estimate of entropies, assuming only a smoothness of the underlying PDF. It is however problematic to simply apply a nearest neighbor estimator (for example the Kozachenko-Leonenko estimator [72]) to each term appearing in eq. 8. This is because the dimensionality of each space associated with the terms differs largely over terms. Thus, a fixed number of neighbors for the search would lead to very different spatial scales (range of distances) for each term. Since the error bias of each term is dependent on these scales, the errors would not cancel each other but accumulate. We therefore use a modified KSG estimator which handles this problem by only fixing the number of neighbors $k$ in the highest dimensional space (k-nearest neighbor search, kNNS) and by projecting the resulting distances to the lower dimensional spaces as the range to look for and count neighbors there (range search, RS) (see [57], type 1 estimator, and [56,74]). In the ensemble variant of TE estimation we proceed by searching for nearest neighbors across points from all repetitions instead of searching the same repetition as the point of reference of the search – thus we form an *ensemble search space* by combining points over repetitions. Finally, the ensemble estimator of TE reads

$$TE_{SPO}(X \to Y, t, u) = \psi(k) + \left\langle \psi\left(n_{\mathbf{y}_{t-1}^{d_Y}(r)} + 1\right) \right. \\ \left. - \psi\left(n_{y_t(r)\,\mathbf{y}_{t-1}^{d_Y}(r)} + 1\right) \right. \\ \left. - \psi\left(n_{\mathbf{y}_{t-1}^{d_Y}(r)\,\mathbf{x}_{t-u}^{d_X}(r)} + 1\right) \right\rangle_r, \tag{9}$$

where $\psi$ denotes the digamma function and the angle brackets $(\langle \cdot \rangle_r)$ indicate an averaging over points in different repetitions $r$ at time instant $t$. The distances to the $k$-th nearest neighbor in the highest dimensional space (spanned by $Y_t, \mathbf{Y}_{t-1}^{d_Y}, \mathbf{X}_{t-u}^{d_X}$) define the radius of the spheres for the counting of the number of points ($n_\cdot$) in these spheres around each state vector ($\cdot$) involved.

In cases where the number of repetitions is not sufficient to provide the necessary amount of data to reliably estimate Shannon entropies through an ensemble average, one may combine ensemble evaluation with collecting realizations over time. In these cases, we count neighbors in a time window $t' \in [t^-, t^+]$ with $t^- \le t' \le t^+$, where $\Delta_t = t^+ - t^-$ controls the temporal resolution of the TE estimation:

$$TE_{SPO}(X \to Y, t', u) = \psi(k) + \left\langle \psi\left(n_{\mathbf{y}_{t'-1}^{d_Y}(r)} + 1\right) \right. \\ \left. - \psi\left(n_{y_{t'}(r)\,\mathbf{y}_{t'-1}^{d_Y}(r)} + 1\right) \right. \\ \left. - \psi\left(n_{\mathbf{y}_{t'-1}^{d_Y}(r)\,\mathbf{x}_{t'-u}^{d_X}(r)} + 1\right) \right\rangle_{r,t'}. \tag{10}$$

**A**

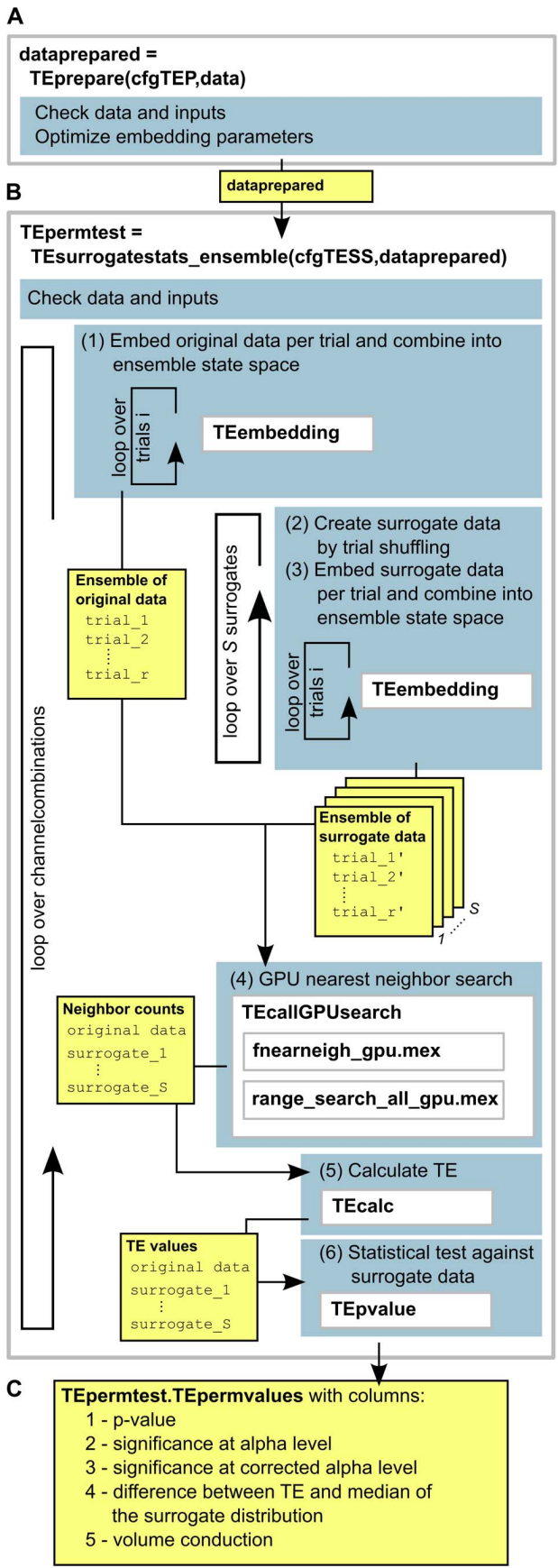

**Figure 2. Transfer entropy estimation using the ensemble method in TRENTOOL 3.0.** (A) Data preparation and optimization of embedding parameters in function `TEprepare.m` ; (B) transfer entropy (TE) estimation from prepared data in `TEsurrogatestats_ensemble.m` (yellow boxes indicate variables being passed between sub-functions). TE is estimated via iterating over all channel combinations provided in the data. For each channel combination: (1) Data is embedded individually per repetition and combined over repetitions into one ensemble state space (chunk), (2) $S$ surrogate data sets are created by shuffling the repetitions of the target time series, (3) each surrogate data set is embedded per repetition and combined into one chunk (forming $S$ chunks in total), (4) $S+1$ chunks of original and surrogate data are passed to the GPU where nearest neighbor searches are conducted in parallel, (5) calculation of TE values from returned neighbor counts for original data and $S$ surrogate data sets using the KSG-estimator [57], (6) statistical testing of original TE value against distribution of surrogate TE values; (C) output of `TEsurrogatestats_ensemble.m`, an array with dimension [no. channels $\times$ 5], where rows hold results for all channel combinations: (1) p-value of TE for this channel combination, (2) significance at the designated alpha level (1 - significant, 0 - not significant), (3) significance after correction for multiple comparisons, (4) absolute difference between the TE value for original data and the median of surrogate TE values, (5) presence of volume conduction (this is always set to 0 when using the ensemble method as instantaneous mixing is by default controlled for by conditioning on the current state of the source time series $x_t(r)$ [119]).

doi:10.1371/journal.pone.0102833.g002

## Implementation

The estimation of TE from finite time series consists of the estimation of joint and marginal entropies as shown in equations 9 and 10, calculated from nearest neighbor statistics, i.e. distances and the count of neighbors within these distances. In practice we obtain these neighbor counts by applying kNNS and RS to reconstructed state spaces. In particular, we use a kNNS in the highest dimensional space to determine the k-th nearest neighbor of a data point and the associated distance. This distance is then used as the range for the RS in the marginal spaces, that return the point counts $n$. Both searches have a high computational cost. This cost increases even further in a practical setting, where we need to calculate TE for a sufficient number of surrogate data sets for statistical testing (see [19] and below for details). To enable TE estimation and statistical testing despite its computational cost, we implemented ad-hoc kNNS and RS algorithms in NVIDIA® CUDA™ C/C++ code [75]. This allows to run thousands of searches in parallel on a modern GPU.

To allow for a better understanding of the parallelization used, we will now briefly describe the main work flow of TE analysis in the open source MathWorks® MATLAB® toolbox TRENTOOL [56], which implements the approach to TE estimation described in the *Background* section. The work flow includes the steps of data preprocessing prior to the use of the GPU algorithm for neighbor searches as well as the statistical testing of resulting TE values. In a subsequent section we will describe the core implementation of the algorithm in more detail and present its integration into TRENTOOL.

## Main analysis work flow in TRENTOOL

**Practical TE estimation in TRENTOOL.** The practical GPU-based TE estimation in TRENTOOL 3.0 is divided into the two steps of data preparation and TE estimation (see Figure 2 and the TRENTOOL 3.0 manual: http: www.trentool.de). As a first step, data is prepared by optimizing embedding parameters for state space reconstruction (Figure 2, panel **A**). As a second step, TE is estimated by following the approach for ensemble-based TE estimation lined out in the preceding section (Figure 2, panel **B**).

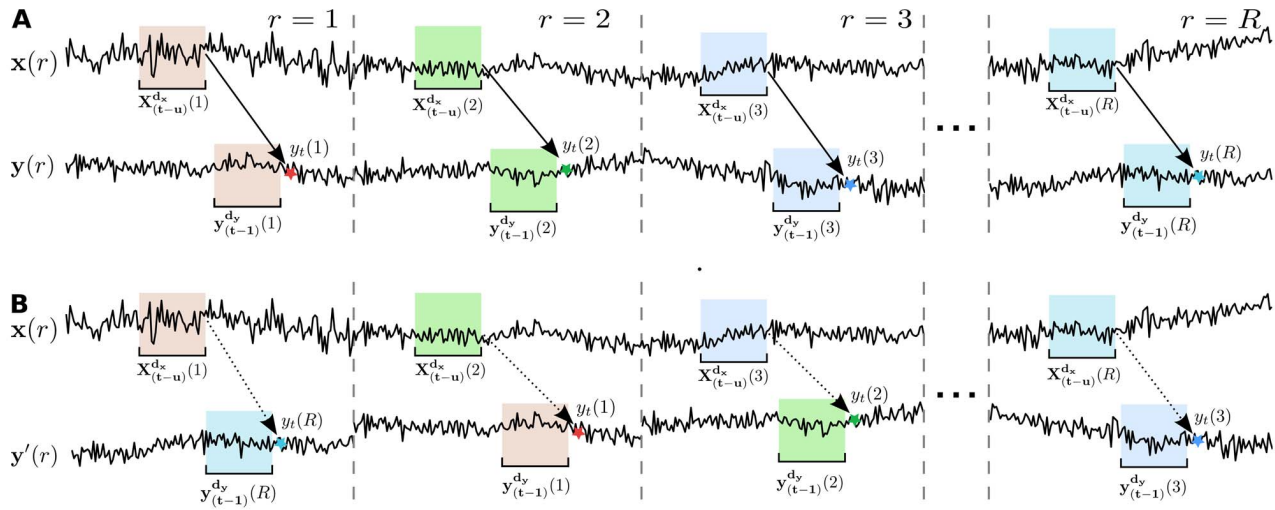

**Figure 3. Creation of surrogate data sets.** (A) Original time series with information transfer (solid arrow) from a source state $\mathbf{x}_{(t-u)}^{d_x}(r)$ to a corresponding target time point $y_t(r)$, given the time point's history $\mathbf{y}_{(t-1)}^{d_y}(r)$. Solid arrows indicate the direction of transfer entropy (TE) analysis, while information transfer is present. (B) Shuffled target time series, repetitions are permutes, such that $y_t(\phi(r))$ and $\mathbf{y}_{(t-1)}^{d_x}(\phi(r))$, where $\phi$ denotes a random permutation. Dashed arrows indicate the direction of TE analysis, while no more information flow is present.
doi:10.1371/journal.pone.0102833.g003

TRENTOOL estimates $TE_{SPO}(X \to Y, t, u)$ (eq. 4) for a given pair of processes $X$ and $Y$ and given values for $u$ and $t$. For each pair, we call $X$ the source and $Y$ the target process.

After data preparation $TE_{SPO}(X \to Y, t, u)$ (eq. 9 and 10) is estimated in six steps: (1) using optimized embedding parameters, original data is embedded per repetition and repetitions are concatenated forming the ensemble search space of the original data, (2) $S$ sets of surrogate data are created from the original data by shuffling the repetitions of the target process $Y$, (3) each surrogate dataset is embedded per repetition and concatenated forming $S$ additional ensemble search spaces for surrogate data, (4) all $S+1$ search spaces of embedded original and surrogate data are passed to a wrapper function that calls the GPU functions to perform individual neighbor searches for each search space in parallel (in the following, we will refer to each of the $S+1$ ensembles as one data *chunk*), (5) TE values are calculated for original and surrogate data chunks from the neighbor counts using the KSG- estimator [57], (6) TE values for original data are tested statistically against the distribution of surrogate TE values.

The proposed GPU algorithm is accessed in step (4). As we will further explain below (see paragraph on *Input data*), the GPU implementation uses the fact that all of the necessary computations on surrogate data sets and the original data are independent and can thus be performed in parallel.

**TE calculation and statistical testing against surrogate data.** Estimated TE values need to be tested for their statistical significance [56] (step (6) of the main TRENTOOL work flow). For this statistical test under a null hypothesis of *no* information transfer between a source $X$ and target time series $Y$, we estimate $TE_{SPO}(X \to Y, t, u)$ and compare it to a distribution of TE values calculated from surrogate data sets. Surrogate data sets are formed by shuffling repetitions in $Y$ to obtain $Y'$, such that $\mathbf{y}_{t}^{d_Y}(r) \to \mathbf{y}_{t}^{d_Y}(\phi(r))$ and $y_t(r) \to y_t(\phi(r))$, where $\phi$ denotes a random permutation of the repetitions $r$ (Figure 3). From this surrogate data set, we calculate surrogate TE values $TE_{SPO}(X \to Y', t, u)$. By repeating this process a sufficient number of times $S$, we obtain a distribution of values $TE_{SPO}(X \to Y', t, u)$. To asses the statistical significance of $TE_{SPO}(X \to Y, t, u)$, we calculate a p-value as the proportion of surrogate TE values $TE_{SPO}(X \to Y', t, u)$ equal or larger than $TE_{SPO}(X \to Y, t, u)$. This p-value is then compared to a critical alpha level (see for example [56,76]).

**Reconstruction of information transfer delays.** $TE_{SPO}(X \to Y, t, u)$ may be used to reconstruct the interaction transfer delay $\delta_{XY}$ between $X$ and $Y$ (eq. 5, [53]). $\delta_{XY}$ may be reconstructed by *scanning* possible values for $u$: $TE_{SPO}(X \to Y, t, u)$ is estimated for all values in $u$; The value that maximizes the $TE_{SPO}(X \to Y, t, u)$ is kept as the reconstructed information transfer delay. We used the reconstruction of information transfer delays as an additional parameter when testing the proposed implementation for correctness and robustness.

## Implementation of the GPU algorithm

**Parallelized nearest neighbor searches.** The KSG estimator used for estimating $TE_{SPO}(X \to Y, t, u)$ in eq. 9 and 10 uses neighbor (distance-)statistics obtained from kNNS and RS algorithms to estimate Shannon differential entropies. Thus, the choice of computationally efficient kNNS and RS algorithms is crucial to any practical implementation of the $TE_{SPO}$ estimator. kNNS algorithms typically return a list of the $k$ nearest neighbors for each reference point, while RS algorithms typically return a list of all neighbors within a given range for each reference point. kNNS and RS algorithms have been studied extensively because of their broad potential for application in nearest neighbor searches and related problems. Several approaches have been proposed to reduce their high computational cost: partitioning of input data into k-d Trees, Quadtrees or equivalent data structures [77] or approximation algorithms (ANN: Approximate Nearest Neighbors) [78,79]. Furthermore, some authors have explored how to parallelize the kNNS algorithm on a GPU using different implementations: exhaustive brute force searches [80,81], tree-based searches [82,83] and ANN searches [83,84].

Although performance of existing implementations of kNNS for GPU was promising, they were not applicable to TE estimation. The most critical reason was that existing implementations did not allow for the concurrent treatment of several problem instances by the GPU and maximum performance was only achieved for very large kNNS problem instances. Unfortunately, the problem instances typically expected in our application are numerous (i.e. $S+1$ problem instances per pair of time series), but rather small compared to the main memory on a typical GPU device in use today. Thus, an implementation that handled only one instance at a time would not have made optimal use of the underlying hardware. Therefore, we designed an implementation that is able to handle several problem instances at once to perform neighbor searches for chunks of embedded original and surrogate data in parallel. Moreover, we aimed at a flexible GPU implementation of kNNS and RS that maximized the use of the GPU's hardware resources for variable configurations of data – thus making the implementation independent of the design of the neuroscientific experiment.

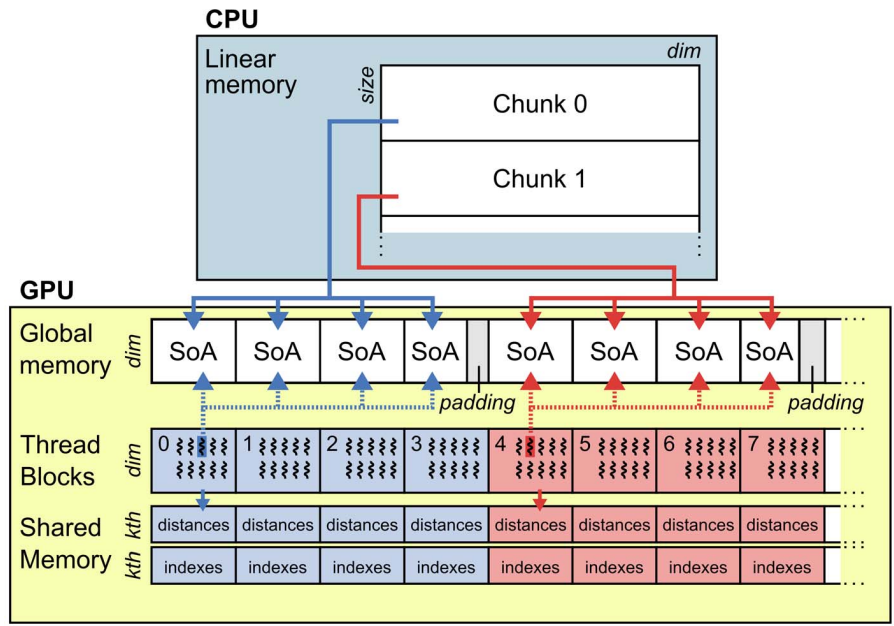

**Figure 4. GPU implementation of the parallelized nearest neighbor search in TRENTOOL 3.0** Chunks of data are prepared on the CPU (embedding and concatenation) and passed to the GPU. Data points are managed in the global memory as Structures of Arrays (SoA). To make maximum use of the memory bandwidth, data is padded to ensure coalesced reading and writing from and to the streaming multiprocessor (SM) units. Each SM handles one chunk in one thread block (dashed box). One block conducts brute force neighbor searches for all data points in the chunk and collects results in its shared memory (red and blue arrows and shaded areas). Results are eventually returned to the CPU.
doi:10.1371/journal.pone.0102833.g004

Our implementation is written in CUDA (Compute Unified Device Architecture) [75] (a port to OpenCL[TM] [85] is work in progress). CUDA is a parallel computing framework created by NVIDIA that includes extensions to high level languages such as C/C++, giving access to the native instruction set and memory of the parallel computational elements in CUDA enabled GPUs. Accelerating an algorithm using CUDA includes translating it into data-parallel sequences of operations and then carefully mapping these operations to the underlying resources to get maximum performance [58,59]. To understand the implementation suggested here, we will give a brief explanation of these resources, i.e. the GPU's hardware architecture, before explaining the implementation in more detail (additionally, see [58,59,75]).

**GPU resources.** GPU resources comprise of massively parallel processors with up to thousands of cores (processing units). These cores are divided among Stream Multiprocessors (SMs) in order to guarantee automatic scalability of the algorithms to different versions of the hardware. Each SM contains 32 to 192 cores that execute operations described in the CUDA kernel code. Operations executed by one core are called a CUDA thread.

Threads are grouped in blocks, which are in turn organized in a grid. The grid is the entry point to the GPU resources. It handles one kernel call at a time and executes it on multiple data in parallel. Within the grid, each block of threads is executed by one SM. The SM executes the threads of a block by issuing them in groups of 32 threads, called warps. Threads within one warp are executed concurrently, while as many warps as possible are scheduled per SM to be resident at a time, such that the utilization of all the cores is maximized.

**Input data.** As input, the proposed RS and kNNS algorithms expect a set of data points representing the search space and a second set of data points that serve as reference points in the searches. One such problem instance is considered one data chunk. Our implementation is able to handle several data chunks simultaneously to make maximum use of the GPU resources. Thus, several chunks may be combined, using an additional index vector to encode the sizes of individual chunks. These chunks are then passed at once to the GPU algorithm to be searched in parallel.

In the estimation of $TE_{SPO}(X \rightarrow Y, t, u)$, according to the work flow described in paragraph *Practical TE estimation in TRENTOOL*, we used the proposed implementation to parallelize neighbor searches over surrogate data sets for a given pair of time series $\mathbf{x}$ and $\mathbf{y}$ and given values for $u$ and $t$. Thus, in one call to the GPU algorithms $S+1$ data chunks were passed as input, where chunks represented the search space for the original pair of time series and $S$ search spaces for corresponding surrogate data sets. Points within the search spaces may have either been collected through temporal or ensemble pooling of embedded data points or a combination of both (eq. 9 or 10).

**Core algorithm.** In the core GPU-based search algorithm, the kNNS implementation is mapped to CUDA threads as depicted in Figure 4 (the RS implementation behaves similarly). Each chunk consists of a set of data points that represents the search space and are at the same time used as reference points for individual searches. Each individual search is handled by one CUDA thread. Parallelization of these searches on the GPU happens in two ways: (1) the GPU algorithm is able to handle several chunks, (2) each chunk can be searched in parallel, such that individual searches within one chunk are handled simultaneously. An individual search is conducted by a CUDA thread by brute-force measuring the infinity norm distance of the given reference point to any other point within the same chunk. Simultaneously, other threads measure these distances for other points in the same chunk or handle a different chunk altogether. Searching several chunks in parallel is an essential feature of the proposed solution, that maximizes the utilization of GPU resources. From the GPU execution point of view, simultaneous searches are realized by handling a variable number of kNNS (or RS) problem instances through one grid launch. The number of searches that can be executed in parallel is thus only limited by the device's global memory that holds the input data and the number of threads that can be started simultaneously (both limitations are taken into account). Furthermore, the solution is implemented such that optimal performance is guaranteed.

**Low-level implementation details.** There are several strategies that are essential for optimal performance when implementing algorithms for GPU devices. Most important are the reduction of memory latencies and the optimal use of hardware resources by ensuring high occupancy (the ratio of number of active warps per SM to the maximum number of possible active warps [58]). To maximize occupancy, we designed our algorithm's kernels such that always more than one block of threads (ideally many) are loaded per SM [58]. We can do this since many searches are executed concurrently in every kernel launch. By maximizing occupancy, we both ensure hardware utilization and improve performance by hiding data memory latency from the GPU's global memory to the SMs' registers [75]. Moreover, in order to reduce memory latencies we take care of input data memory alignment and guarantee that memory readings issued by the threads of a warp are coalesced into as few memory transfers as possible. Additionally, with the aim of minimizing sparse data accesses to memory, data points are organized as Structures of Arrays (SoA). Finally, we use the shared memory inside the SMs (a self-programmed intermediate cache between global memory and SMs) to keep track of nearest neighbors associated information during searches. The amount of shared memory and registers is limited in a SM. The maximum possible occupancy depends on the number of registers and shared memory needed by a block, which in turn depends on the number of threads in the block. For our implementation, we used a suitable block size of 512 threads.

**Implementation interface.** The GPU functionality is accessed through MATLAB scripts for kNNS ('fnearneigh_gpu.mex') and RS ('range_search_all_gpu.mex'), which encapsulate all the associated complexity. Both scripts are called from TRENTOOL using a wrapper function. In its current implementation in TRENTOOL (see paragraph *Practical TE estimation in TRENTOOL*), the wrapper function takes all $S+1$ chunks as input and launches a kernel that searches all chunks in parallel through the mex-files for kNNS and RS. The wrapper makes sure that the input size does not exceed the GPU device's available global memory and the maximum number of threads that can be started simultaneously. If necessary, the wrapper function splits the input into several kernel calls; it also manages the output, i.e. the neighbor counts for each chunk, which are passed on for TE calculation.

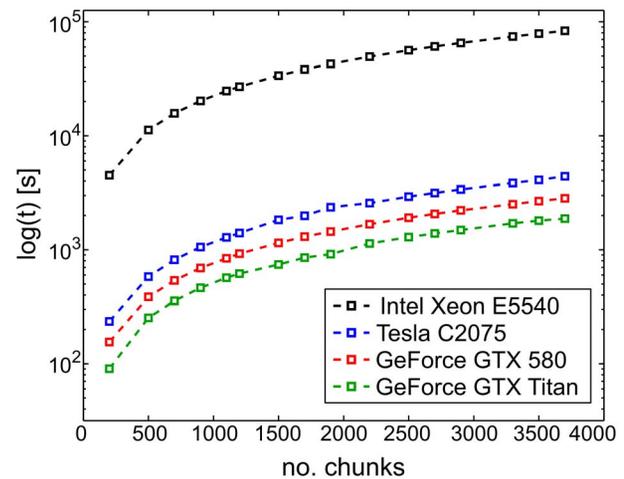

**Figure 5. Practical performance measures of the ensemble method for GPU compared to CPU.** Combined execution times in s for serial and parallel implementations of k-nearest neighbor and range search as a function of input size (number of data chunks). Execution times were measured for the serial implementation running on a CPU (black) and for our parallel implementation using one of three GPU devices (blue, red, green) of varying computing power. Computation using a GPU was considerably faster than using a CPU (by factors 22, 33 and 50 respectively).
doi:10.1371/journal.pone.0102833.g005

## Evaluation

To evaluate the proposed algorithm we investigated four properties: first, whether the speedup is sufficient to allow the application of the method to real-world neural datasets; second, the correctness of results on simulated data, where the ground truth is known; third, the robustness of the algorithm for limited sample sizes; fourth, whether plausible results are achieved on a neural example dataset.

### Ethics statement

The neural example dataset was taken from an experiment described in [86]. All subjects gave written informed consent before the experiment. The study was approved by the local ethics committee (Johann Wolfgang Goethe University, Frankfurt, Germany).

### Evaluation of computational speedup

To test for an increase in performance due to the parallelization of neighbor searches, we compared practical execution times of the proposed GPU implementation to execution times of the serial kNNS and RS algorithms implemented in the MATLAB toolbox TSTOOL (http: www.dpi.physik.uni-goettingen.de/tstool/). This toolbox wraps a FORTRAN implementation of kNNS and RS, and has proven the fastest CPU toolbox for our purpose. All testing was done in MATLAB 2008b (MATLAB 7.7, The MathWorks Inc., Natick, MA, 2008). As input, we used increasing numbers of chunks of simulated data from two coupled Lorenz systems, further described below. Repetitions of simulated time series were embedded and combined to form ensemble state spaces, i.e. chunks of data (c.f. paragraph *Input Data*). To obtain increasing input sizes, we duplicated these chunks the desired number of times. While the CPU implementation needed to iteratively perform searches on individual chunks, the GPU implementation searched chunks in parallel (note that chunks are treated independently here, so that there is no speedup because of the duplicated chunk data). Note that for both, CPU and GPU implementations, data handling prior to nearest neighbor searches is identical. We were thus able to confine the testing of performance differences to the respective kNNS and RS algorithms only, as all data handling prior to nearest neighbor searches was conducted using the same, highly optimized TRENTOOL functionalities.

Analogous to TE estimation implemented in TRENTOOL, we conducted one kNNS (with $k = 4$, TRENTOOL default, see also [87]) in the highest dimensional space and used the returned distances for a RS in one lower dimensional space. Both functions were called for increasing numbers of chunks to obtain the execution time as a function of input size. One chunk of data from the highest dimensional space had dimensions $[30094 \times 17]$ and size 1.952 MB (single precision); one chunk of data from the lower dimensional space had dimensions $[30094 \times 8]$ and size 0.918 MB (single precision). Performance testing of the serial implementation was carried out on an Intel Xeon CPU (E5540, clocked at 2.53 GHz), where we measured execution times of the TSTOOL kNNS (functions 'nn_prepare.m' and 'nn_search.m') and the TSTOOL RS (function 'range_search.m'). Testing of the parallel implementation was carried out on GPU devices of varying processing power (NVIDIA Tesla C2075, GeForce GTX 580 and GeForce GTX Titan). On the GPUs, we measured execution times for the proposed kNNS ('fnearneigh_gpu.mex') and RS ('range_search_all_gpu.mex') implementation. When the GPU's global memory capacity was exceeded by higher input sizes, data was split and computed over several runs (i.e. calls to the GPU). All performance testing was done by measuring execution times using the MATLAB functions tic and toc.

To obtain reliable results for the serial implementation we ran both kNNS and RS 200 times on the data, receiving an average execution time of 1.26 s for kNNS and an average execution time of 24.1 s for RS. We extrapolated these execution times to higher numbers of chunks and compared them to measured execution times of the parallel searches on three NVIDIA GPU devices. On average, execution times on the GPU compared to the CPU were faster by a factor of 22 on the NVIDIA Tesla C2075, by a factor of 33 for the NVIDIA GTX 580 and by a factor of 50 for the NVIDIA GTX Titan (Figure 5).

To put these numbers into perspective, we note that in a neuroscience experiment the number of chunks to be processed is the product of (typical numbers): channel pairs for TE (100) * number of surrogate data sets (1000) * experimental conditions (4) * number of subjects (15). This results in a total computational load on the order of $6 * 10^6$ chunks to be processed. Given an execution time of 24.1 s/50 on the NVIDIA GTX Titan for a typical test dataset, these computations will take $2.9 * 10^6 s$ or 4.8 weeks on a single GPU, which is feasible compared to the initial duration of 240 weeks on a single CPU. Even when considering a trivial parallelization of the computations over multiple CPU cores and CPUs, the GPU based solution is by far more cost and energy efficient than any possible CPU-based solution. If in addition a scanning of various possible information transfer delays is important, then parallelization over multiple GPUs seems to be the only viable option.

## Evaluation on Lorenz systems

To test the ability of the presented implementation to successfully reconstruct information transfer between systems with a non-stationary coupling, we simulated various coupling scenarios between stochastic and deterministic systems. We introduced non-stationary into the coupling of two processes by varying the coupling strength over the course of a repetition (all other parameters were held constant). Simulations for individual scenarios are described in detail below. For the estimation of TE we used MathWork's MATLAB, and the TRENTOOL toolbox extended by the implementation of the ensemble method proposed above (version 3.0, see also [56] and http: www.trentool.de). For a detailed testing of the used estimator $TE_{SPO}$ (eq. 4) refer to [53].

**Coupled Lorenz systems.** Simulated data was taken from two unidirectionally coupled Lorenz systems labeled $X$ and $Y$. Systems interacted in direction $X \to Y$ according to equations:

$$\dot{U}_i(t) = \sigma(V_i(t) - U_i(t)),$$

$$\dot{V}_i(t) = U_i(t)(\rho_i - W_i(t)) - V_i(t) + \sum_{i,j = X,Y} \gamma_{ij} V_j^2(t - \delta_{ij}), \quad (11)$$

$$\dot{W}_i(t) = U_i(t)V_i(t) - \beta W_i(t),$$

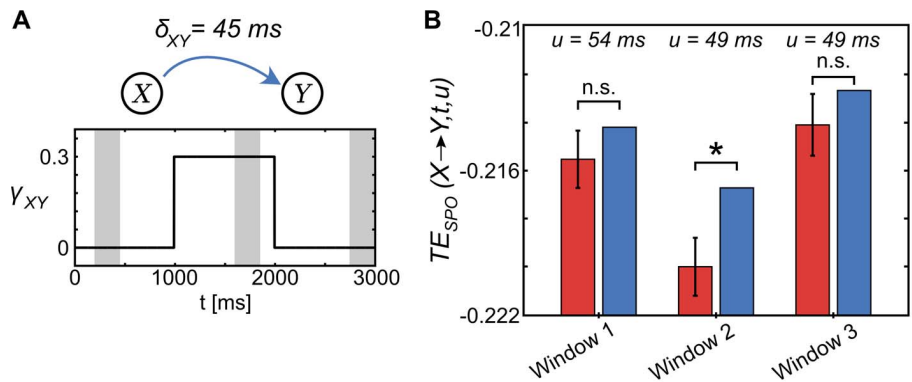

**Figure 6. Transfer entropy reconstruction from non-stationary Lorenz systems.** We used two dynamically coupled Lorenz systems (A) to simulate non-stationarity in data generating processes. A coupling $\gamma_{XY} = 0.3$ was present during a time interval from 1000 to 2000 ms only ($\gamma_{XY} = 0$ otherwise). The information transfer delay was set to $\delta_{XY} = 45 ms$. Transfer entropy (TE) values were reconstructed using the ensemble method combined with the scanning approach proposed in [53] to reconstruct information transfer delays. Assumed delays $u$ were scanned from 35 to 55 ms (1 ms resolution). In (B) the maximum TE values for original data over this interval are shown in blue. Red bars indicate the corresponding mean over surrogate TE values (error bars indicate 1 SD). Significant TE was found for the second time window only; here, the delay was reconstructed as $u = 49 ms$.
doi:10.1371/journal.pone.0102833.g006

**Table 1.** Parameter settings for simulated autoregressive processes.

| Testcase | $\alpha_X$ | $\alpha_Y$ | $\beta_{YX}$ | $\beta_{XY}$ | $\delta_{YX}$ | $\delta_{XY}$ |
|---|---|---|---|---|---|---|
| Unidirectional | 0.75 | 0.35 | 0 | −0.35 | 0 | 10 |
| Two-step unidirectional | 0.75 | 0.35 | 0 | −0.35 | 0 | 10 |
| Bidirectional | 0.475 | 0.35 | −0.4 | −0.35 | 20 | 10 |

where $i,j = X, Y$, $\delta_{ij}$ is the coupling delay and $\gamma_{ij}$ is the coupling strength; $\sigma$, $\rho$ and $\beta$ are the *Prandtl number*, the *Rayleigh number*, and a geometrical scale. Note, that $\gamma_{YX} = \gamma_{XX} = \gamma_{YY} = 0$ for the test cases (no self feedback, no coupling from $Y$ to $X$). Numerical solutions to these differential equations were computed using the *dde23* solver in MATLAB and results were resampled such that the delays amounted to the values given below. For analysis purposes we analyzed the V-coordinates of the systems.

We introduced non-stationarity in the coupling between both systems by varying the coupling strength $\gamma$ over time. In particular, a coupling $\gamma_{XY} = 0.3$ was set for a limited time interval only, whereas before and after the coupling interval $\gamma_{XY}$ was set to 0. A constant information transfer delay $\delta_{XY} = 45ms$ was simulated for the whole coupling interval. We simulated 150 repetitions with 3000 data points each, with a coupling interval from approximately 1000 to 2000 data points (see Figure 6, panel **A**).

For each scenario, 500 surrogate data sets were computed to allow for statistical testing of the reconstructed information transfer. Surrogate data were created by permutation of data points in blocks of the target time series (Figure 3), leaving each repetition intact. The value $k$ for the nearest neighbor search was set to 4 for all analyses (TRENTOOL default, see also [87]).

**Results.** We analyzed data from three time windows from 200 to 450 ms, 1600 to 1850 ms and 2750 to 3000 ms using the estimator proposed in eq. 10 with $\Delta_t = 250ms$, assuming local stationarity (Figure 6, panel **A**). For each time window, we scanned assumed delays in the interval $u = [35, 55]$. Figure 6, panel **B**, shows the maximum TE value from original data (blue) over all assumed $u$ and the corresponding mean surrogate TE value (red). Significant differences between original TE and surrogate TE were found in the second time window only (indicated by an asterisk). No significant information transfer was found during the non-coupling intervals. The information transfer delay reconstructed for the second analysis window was 49 ms (true information transfer delay $\delta_{XY} = 45ms$). Thus, the proposed implementation was able to reliably detect a coupling between both systems and reconstructed the corresponding information transfer delay with an error of less than 10%.

## Evaluation on autoregressive processes

To asses the performance of the proposed implementation on non-abrupt changes in coupling, we simulated various coupling scenarios for two autoregressive processes $X$, $Y$ of order 1 (AR(1)-processes) with variable couplings over time. In each scenario, couplings were modulated using hyperbolic functions to realize a smooth transition between uncoupled and coupled regimes. The AR(1)-processes were simulated according to the equations

$$x(t) = \alpha_X x(t-1) + \gamma_{YX}(t) y(t - \delta_{YX}) + \eta_X(t), \quad (12)$$

$$y(t) = \alpha_Y y(t-1) + \gamma_{XY}(t) x(t - \delta_{XY}) + \eta_Y(t), \quad (13)$$

where $\alpha_X$, $\alpha_Y$ are the AR parameters, $\gamma_{YX}(t)$, $\gamma_{XY}(t)$ denote coupling strength, $\delta_{YX}$, $\delta_{XY}$ are the coupling delays and $\eta_X$, $\eta_Y$ denote uncorrelated, unit-variance, zero-mean Gaussian white noise terms.

**Simulated coupling scenarios.** We simulated three coupling scenarios, where the coupling varied in strength over the course of a repetition (duration 3000 ms): (1) unidirectional coupling $X \to Y$ with a coupling onset around 1000 ms; (2) unidirectional coupling with a two-step increase in coupling $X \to Y$ at around 1000 ms and around 2000 ms; (3) bidirectional coupling $X \to Y$ with onset around 1000 ms and $Y \to X$ with onset around 2000 ms. See table 1 for specific parameter values used in each scenario.

We realized a varying coupling strength $\gamma_{XY}(t)$ (and $\gamma_{YX}(t)$ for scenario (3)) by modulating coupling parameters $\beta_{YX}$, $\beta_{XY}$ with a hyperbolic tangent function. No coupling was realized by setting $\beta_{\cdot} = 0$. For scenarios (1) and (3) we used the coupling

$$\gamma_{YX} = \beta_{YX} * 0.5(1 + \tanh[0.05(t - 2000)]) \quad (14)$$

$$\gamma_{XY} = \beta_{XY} * 0.5(1 + \tanh[0.05(t - 1000)]), \quad (15)$$

where 0.05 was the slope and 2000 and 1000 are the inflection points of the hyperbolic tangent respectively. Note that we additionally scaled the tanh function such that function value ranged from 0 to 1. For coupling scenario (2), the two-step increase in $\gamma_{XY}$ was expressed as:

$$\begin{aligned} \gamma_{XY} = \beta_{XY} * 0.5 & [0.5(1 + \tanh[0.05(t - 1000)]) \\ & + 0.5(1 + \tanh[0.05(t - 2000)])] \end{aligned} \quad (16)$$

We chose the arguments of the hyperbolic function such that the function's slope led to a smooth increase in the coupling over an epoch of approximately 200 ms around the inflection points at 1 and 2 s (Figure 7, panels **A–D**). For each scenario, we simulated 50 trials of length 3000 ms with a sampling rate of 1000 Hz. We then estimated time resolved TE for analysis windows of length $\Delta_t = 300ms$. Again, we mixed temporal and ensemble pooling according to eq. 10. For the scenario with unidirectional coupling (1) we used four analysis windows to cover the change in coupling (from 0.2 to 0.5 s, 0.5 to 0.8 s, 0.8 to 1.1 s, and 1.1 to 1.4 s, see Figure 7, panel **E**), for the two-step increase

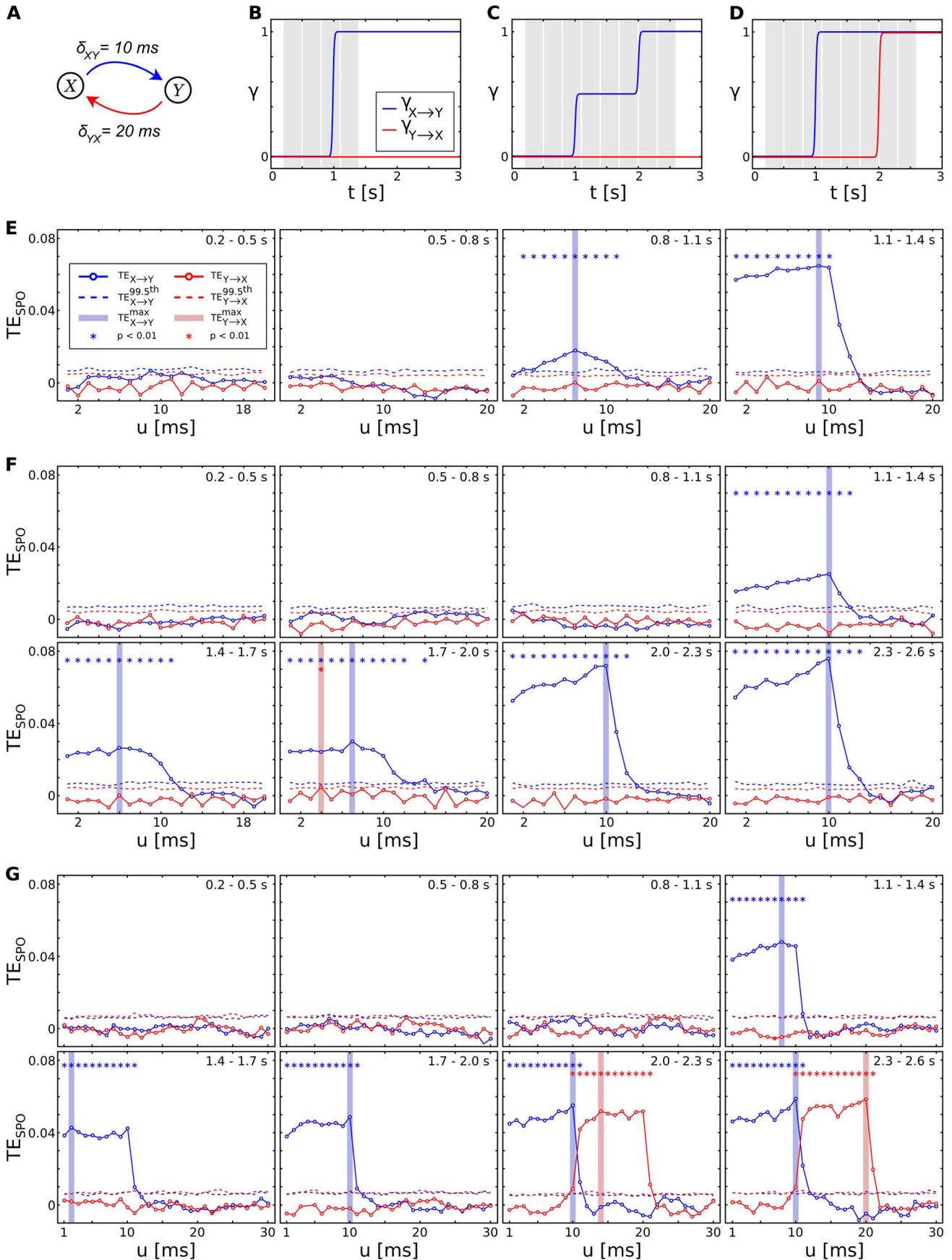

(2) and bidirectional (3) scenarios, we used eight analysis windows each (from 0.2 to 0.5 s, 0.5 to 0.8 s, 0.8 to 1.1 s, 1.1 to 1.4 s, 1.4 to 1.7 s, 1.7 to 2.0 s, 2.0 to 2.3 s, and 2.3 to 2.6 s, see Figure 7, panels **F** and **G**). As for the Lorenz systems, 500 surrogate data sets were used for the statistical testing in each analysis. Surrogate data were created by blockwise (i.e. repetitionwise) permutation of data points in the target time series. The value $k$ for the nearest neighbor search was set to 4 for all analyses (TRENTOOL default, see also [87]).

**Results – Scenario (1), unidirectional coupling.** For scenario (1) of two unidirectionally coupled AR(1)-processes with a delay $\delta_{XY} = 10ms$, we used a scanning approach [53] to reconstruct TE and the corresponding information transfer delay. We scanned assumed delays in the interval $u = [1,20]$ and used four analysis windows of length 300 ms each, ranging from 0.2 to 1.4 s. For the first two analysis windows, no significant information transfer was found (0.2 to 0.5 and 0.5 to 0.8 s). For the third and fourth analysis window we detected significant TE, where we found a maximum significant TE value at 7 ms for the third analysis window (0.8 to 1.1 s) and a maximum at 9 ms for the fourth window (1.1 to 1.4 s). Thus, the proposed implementation was able to detect information transfer between both processes if present (later than 1.1 s). During the transition in coupling strength between 0.8 and 1.1 s TE was detected, but the method showed a small error in the reconstructed information transfer delay. This may be due to too little data to detect the weaker coupling at this epoch of the simulated coupling (see below).

**Results – Scenario (2), unidirectional coupling with two-step increase.** For scenario (2), we again used the scanning approach for TE reconstruction, using an interval of assumed delays $u = [1,20]$, where the true delay was simulated at $\delta_{XY} = 10ms$. No TE was detected prior to the coupling onset around 1 s. TE was detected for analysis windows 4, 5, and 6 (1.1 to 1.4, 1.4 to 1.7, 1.7 to 2.0 s) with reconstructed information transfer delays of 10, 4, and 7 ms respectively. Further, significant TE was found for analysis windows 7 and 8 (after the second increase in coupling strength around 2 s). Here, the correct coupling of 10 ms was reconstructed. One false positive result was obtained in window 6 (1.7 to 2.0 s), where significant TE was found in the direction $Y \to X$.

Note, that the method's ability to recover information transfer from data depends on the strength of the coupling relative to the amount of data that is available for TE estimation. This is observable in the reconstructed TE in the third analysis window for scenario (1) and (2): in scenario (2) no TE is detected, whereas in scenario (1) weak information transfer is already reconstructed for the third window. Note, that in scenario (2) the simulated coupling between 1 and 2 s is much weaker than the coupling in

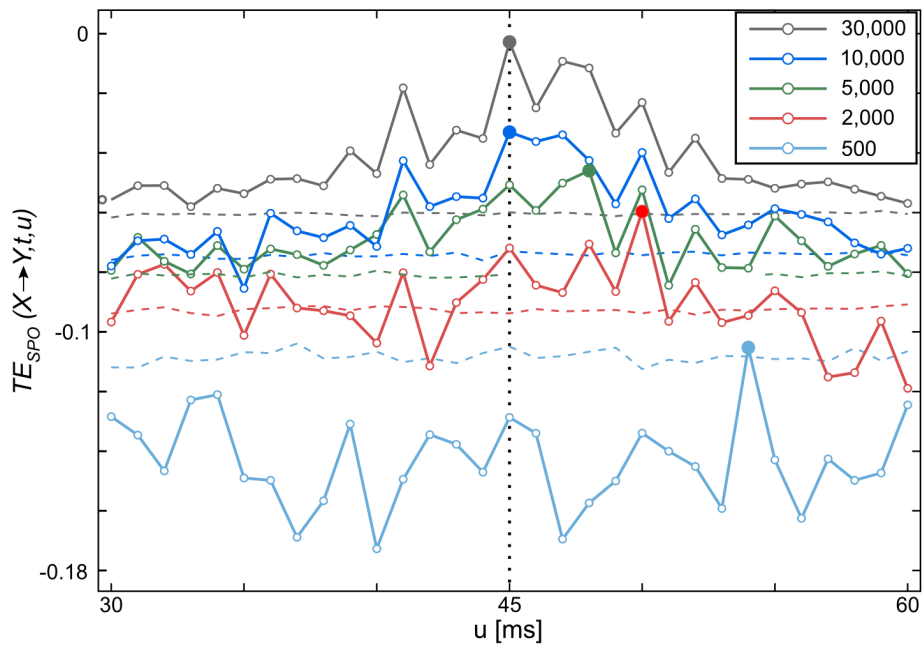

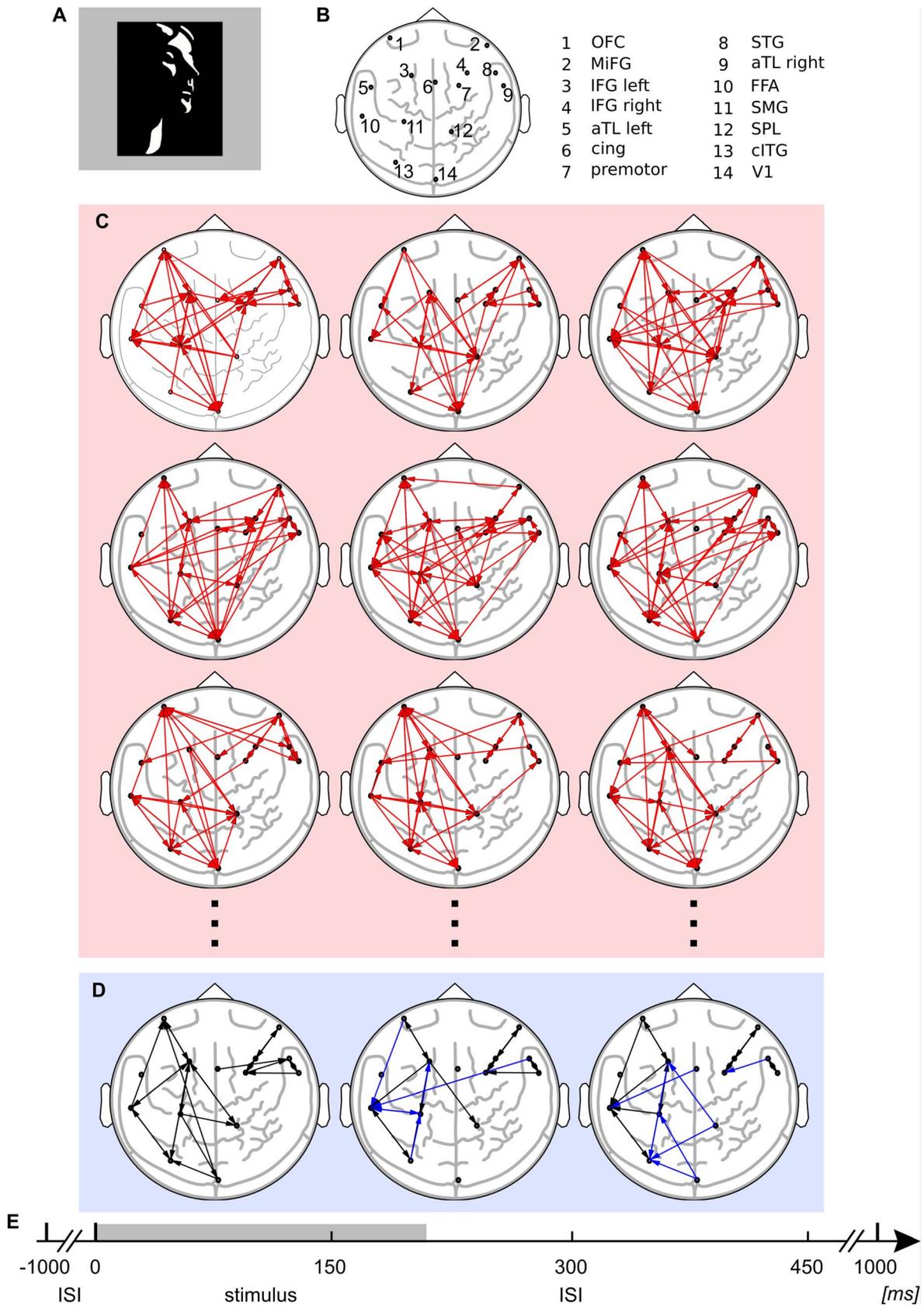

**Figure 9. Transfer entropy reconstruction from electrophysiological data.** Time resolved reconstruction of transfer entropy (TE) from magnetoencephalographic (MEG) source data, recorded during a face recognition task. (A) Face stimulus [88]. (B) Cortical sources after beamforming of MEG data (L, left; R, right: L orbitofrontal cortex (OFC); R middle frontal gyrus (MiFG); L inferior frontal gyrus (IFG left); R inferior frontal gyrus (IFG

right); L anterior inferotemporal cortex (aTL left); L cingulate gyrus (cing); R premotor cortex (premotor); R superior temporal gyrus (STG); R anterior inferotemporal cortex (aTL right); L fusiform gyrus (FFA); L angular/supramarginal gyrus (SMG); R superior parietal lobule/precuneus (SPL); L caudal ITG/LOC (cITG); R primary visual cortex (V1)). (C) Reconstructed TE in three single subjects (red box) in three time windows (0—150 ms, 150—300 ms, 300—450 ms). Each link (red arrows) corresponds to significant TE on single subject level (corrected for multiple comparisons). (D) Thresholded TE links over 15 subjects (blue box) in three time windows (0—150 ms, 150—300 ms, 300—450 ms). Each link (black arrows) corresponds to significant TE in eight and more individual subjects ($p < <0.0001^{***}$, after correction for multiple comparisons). Blue arrows indicate differences between time windows, i.e. links that occur for the first time in the respective window. (E) Experimental design: stimulus was presented for 200 ms (gray shading), during the inter stimulus interval (ISI, 1800 ms) a fixation cross was displayed.
doi:10.1371/journal.pone.0102833.g009

the unidirectional scenario (1) (Figure 7, panels **C** and **B**). This resulted in smaller and non-significant absolute TE values and in reconstructed information transfer delays that were less precise.

**Results – Scenario (3), bidirectional coupling.** For scenario (3), we used the scanning approach for TE reconstruction, using an interval of assumed delays $u=[1,30]$, where the true delay was simulated at $\delta_{XY}=10ms$ and $\delta_{YX}=20ms$. No TE in either direction was detected prior to the first coupling onset around 1 s. TE for the first direction $X \to Y$ was detected after coupling onset around 1 s for analysis windows 4, 5, 6, 7, and 8. Reconstructed information transfer delays were 8 and 2 ms for analysis windows 4 and 5. For each of the following analysis windows 6 to 8 the correct delay of 10 ms was reconstructed.

TE for the second direction $Y \to X$ was detected after coupling onset around 2 s for analysis windows 7 and 8, where also the correct coupling of 20 ms was reconstructed. Thus, the proposed implementation was able to reconstruct information transfer in bidirectionally coupled systems.

## Evaluation of the robustness of ensemble-based TE-estimation

We tested the robustness of the ensemble method for cases where the amount of data available for TE estimation was severely limited. We created two coupled Lorenz systems $X$, $Y$ from which we sampled a maximum number of 300 repetitions of 300 ms each at 1000 Hz, using a coupling delay of $\delta_{XY}=45ms$ (see equation 11). We embedded the resulting data with their optimal embedding parameters for different values of the assumed delay $u$ (30 to 60 ms, step size of 1 ms, also see equation 4). From the embedded data, we used subsets of data points with varying size $M$ ($M=\{500,2000,5000,10000,30000\}$) to estimate TE according to equation 10 (we always used the first $M$ consecutive data points for TE estimation). For each $u$ and number of data points $M$, we created surrogate data to test the estimated TE value for statistical significance. Furthermore, we reconstructed the corresponding information transfer delay for each $M$ by finding the maximum TE value over all values for $u$. A reconstructed TE value was considered a robust estimation of the simulated coupling if the reconstructed delay value was able to recover the simulated information transfer delay of $45ms$ with an error of $\pm 5\%$, i.e. $45 \pm 1.125ms$.

A sufficiently accurate reconstruction was reached for 10000 and 30000 data points (Figure 8). For 5000 data points estimation was off by approximately 7% (the reconstructed information transfer delay was 48 ms), less data entering the estimation led to a further decline in accuracy of the recovered information transfer delay (here, reconstructed delays were 50 ms and 54 ms for 2000 and 500 data points respectively).

## Evaluation on neural time series from magnetoencephalography

To demonstrate the proposed method's suitability for time-resolved reconstruction of information transfer and the corresponding delays from biological time series, we analyzed magnetoencephalographic (MEG) recordings from a perceptual closure experiment described in [86].

**Subjects.** MEG data were obtained from 15 healthy subjects (11 females; mean $\pm$ SD age, $25.4 \pm 5.6$ years), recruited from the local community.

**Task.** Subjects were presented with a randomized sequence of degraded black and white picture of human faces [88] (Figure 9, panel **A**) and scrambled stimuli, where black and white patches were randomly rearranged to minimize the likelihood of detecting a face. Subjects had to indicate the detection of a face or no-face by a button press. Each stimulus was presented for 200 ms, with a random inter-repetition interval (IRI) of 3500 to 4500 ms (9, panel **E**). For further analysis we used repetitions with correctly identified face conditions only.

**MEG and MRI data acquisition.** MEG data were recorded using a 275-channel whole-head system (Omega 2005, VSM MedTech Ltd., BC, Canada) at a rate of 600 Hz in a synthetic third order axial gradiometer configuration. The data were filtered with 4th order Butterworth filters with 0.5 Hz high-pass and 150 Hz low-pass. Behavioral responses were recorded using a fiber optic response pad (Lumitouch, Photon Control Inc., Burnaby, BC, Canada).

Structural magnetic resonance images (MRI) were obtained with a 3 T Siemens Allegra, using 3D magnetization-prepared rapid-acquisition gradient echo sequence. Anatomical images were used to create individual head models for MEG source reconstruction.

**Data analysis.** MEG data were analyzed using the open source MATLAB toolboxes FieldTrip (version 2012-12-08; [89], SPM2 ⟨http://www.fil.ion.ucl.ac.uk/spm/⟩, and TRENTOOL [56]. We will briefly describe the applied analysis here, for a more in depth treatment refer to [86].

For data preprocessing, data epochs (repetitions) were defined from the continuously recorded MEG signals from $-1000$ to 1000 ms with respect to the onset of the visual stimulus. Only data repetitions with correct responses were considered for analysis. Data epochs contaminated by eye blinks, muscle activity, or jump artifacts in the sensors were discarded. Data epochs were baseline corrected by subtracting the mean amplitude during an epoch ranging from $-500$ to $-100$ ms before stimulus onset.

To investigate differences in source activation in the face and non-face condition, we used a frequency domain beamformer [90] at frequencies of interest that had been identified at the sensor level (80 Hz with a spectral smoothing of 20 Hz). We computed the frequency domain beamformer filters for combined data epochs ("common filters") consisting of activation (multiple windows, duration, 200 ms; onsets at every 50 ms from 0 to 450 ms) and baseline data ($-350$ to $-150$ ms) for each analysis interval. To compensate for the short duration of the data windows, we used a regularization of $\lambda = 5\%$ [91].

To find significant source activations in the face versus non-face condition, we first conducted a within-subject t-test for activation versus baseline effects. Next, the t-values of this test statistic were subjected to a second-level randomization test at the group level to obtain effects of differences between face and no-face conditions; a

**Table 2.** Reconstructed information transfer delays for magnetoencephalographic data.

| Source | Target | 0–150 ms | 150–300 ms | 300–450 ms |
|--------|--------|----------|------------|------------|
| SPL | IFG left | 5.00 | - | 5.50 |
| SPL | cITG | - | - | 5.00 |
| cITG | IFG left | 5.00 | 5.00 | 5.00 |
| cITG | FFA | 5.00 | 5.00 | 5.00 |
| cITG | SMG | - | 5.00 | - |
| STG | aTL right | 5.00 | 5.00 | 5.00 |
| STG | Premotor | - | - | 5.83 |
| STG | FFA | - | 5.50 | - |
| aTL right | STG | 5.00 | 5.00 | 5.00 |
| aTL right | Premotor | 5.00 | 5.60 | - |
| SMG | SPL | 5.00 | - | - |
| SMG | V1 | 5.00 | - | - |
| SMG | IFG left | - | 5.20 | - |
| SMG | FFA | - | 5.22 | 5.20 |
| OFC | IFG left | 5.18 | 5.00 | 5.00 |
| OFC | FFA | - | 5.00 | 5.20 |
| MiFG | IFG right | 5.00 | 5.00 | 5.00 |
| MiFG | Premotor | 5.00 | 5.00 | 5.00 |
| IFG right | MiFG | 5.00 | 5.00 | 5.00 |
| IFG right | Premotor | 5.00 | 5.00 | 5.00 |
| IFG left | SPL | 5.00 | 5.00 | - |
| IFG left | cITG | 5.00 | - | 5.00 |
| IFG left | SMG | 5.00 | 5.40 | 5.00 |
| IFG left | OFC | 5.00 | 5.00 | 5.00 |
| IFG left | FFA | 5.00 | 5.00 | 5.00 |
| FFA | cITG | 5.00 | 5.00 | 5.22 |
| FFA | OFC | 5.00 | - | - |
| FFA | IFG left | 5.00 | - | - |
| FFA | SMG | - | 5.00 | - |
| V1 | cITG | 5.25 | - | 5.25 |
| V1 | SMG | - | - | 5.00 |
| Premotor | STG | 5.00 | - | - |
| Premotor | MiFG | 5.00 | 5.00 | 5.00 |
| Premotor | IFG right | 5.00 | 5.00 | 5.00 |
| Cing | STG | 5.25 | - | - |
| Cing | FFA | - | - | 5.67 |

Mean over reconstructed interaction delays for significant information transfers in three analysis windows. Information transfer delays were investigated in steps of 2 ms, from 5–17 ms. Fractional numbers arise from averaging over subjects.
doi:10.1371/journal.pone.0102833.t002

p-value $<0.01$ was considered significant. We identified 14 sources with differential spectral power between both conditions in the frequency band of interest in occipital, parietal, temporal, and frontal cortices (see Figure 9, panel **B**, and [86] for exact anatomical locations). We then reconstructed source time courses for TE analysis, this time using a broadband beamformer with a bandwidth of 10 to 150 Hz.

We estimated TE between beamformer source time courses using our ensemble method with a mixed pooling of embedded time points over repetitions $r$ and time windows $t'$ (eq. 10). We analyzed three non-overlapping time windows $\Delta_t$ of 150 ms each (0–150 ms, 150–300 ms, 300–450 ms, Figure 9, panel **C**). We furthermore reconstructed information transfer delays for significant information transfer by scanning over a range of assumed delays from 5 to 17 ms (resolution 2 ms), following the approach in [53]. We corrected the resulting information transfer pattern for cascade effects as well as common drive effects using a graph-based post-hoc correction proposed in [54].

**Results.** Time-resolved GPU-based TE analysis revealed significant information transfer at the group-level ($p < < 0.001$ corrected for multiple comparison; binomial test under the null hypothesis of the number of occurrences $k$ of a link being $B(k|p_0,n)$-distributed, where $p_0 = 0.05$ and $n = 15$), that changed over time (Figure 9, panel **D** and table 2 for reconstructed information transfer delays). Our preliminary findings of information transfer are in line with hypothesis formulated in [92], [93] and [86], and the time-dependent changes show the our method's sensitivity to the dynamics of information processing during experimental stimulation, in line with the simulation results above.

## Discussion

### Efficient transfer entropy estimation from an ensemble of time series

We presented an efficient implementation of the ensemble method for TE estimation by [55]. As laid out in the introduction, estimating TE from an ensemble of data allows to analyze information transfer between time series that are non-stationary and enables the estimation of TE in a time-resolved fashion. This is especially relevant to neuroscientific experiments, where rapidly changing (and thus non-stationary) neural activity is believed to reflect neural information processing. However, up until now the ensemble method has remained out of reach for application in neuroscience because of its computational cost. Only with using parallelization on a GPU, as presented here, the ensemble method becomes a viable tool for the analysis of neural data. Thus, our approach makes it possible for the first time to efficiently analyze information transfer between neural time series on short time scales. This allows us to handle the non-stationarity of underlying processes and makes a time- resolved estimation of TE possible. To facilitate the use of the ensemble method it has been implemented as part of the open source toolbox TREN-TOOL (version 3.0).

Even though we will focus on neural data when discussing applications of the ensemble method for TE estimation below, this approach is well suited for applications in other fields. For example, TE as defined in [14] has been applied in physiology [42–44], climatology [94,95], financial time series analysis [45,96], and in the theory of cellular automata [48]. Large datasets from these and other fields may now be easily analyzed with the presented approach and its implementation in TRENTOOL.

### Notes on the practical application of the ensemble method for TE estimation

**Applicability to simulated and real world experimental data.** To validate the proposed implementation of the ensemble method, we applied it to simulated data as well as MEG recordings. For simulated data, information transfer could reliably be reconstructed despite the non-stationarity in the underlying generating processes. For MEG data the obtained speed-up was large enough to analyze these data in practical time. Information transfer reconstructed in a time-resolved fashion from the MEG source data was in line with findings by [86,92,93], as discussed below.

Note, that even though our proposed implementation of the ensemble method reduces analysis times by a significant amount, the estimation of TE from neural time series is still time consuming relative to other measures of connectivity. For the example MEG data set presented in this paper, TE estimation for one subject and one analysis window took 93 hours on average (when scanning over seven values for the assumed information transfer delay $u$ and reconstructing TE for all possible combinations of 14 sources). Thus, for 15 subjects with three analysis windows each, the whole analysis would take approximately six months when carried out in a serial fashion on one computer equipped with a modern GPU (e.g. NVIDIA GTX Titan). This time may however be reduced by parallelizing the analysis over subjects and analysis windows on multiple GPUs, as it was done for this study.

**Available data and choice of window size.** As available data is often limited in neuroscience and other real-world applications, the user has to make sure that enough data enters the analysis, such that a reliable estimation of TE is possible. In the proposed implementation of the ensemble method for TE estimation the amount of data entering the estimation directly depends on the size of the chosen analysis window and the amount of available repetitions of the process being analyzed. Furthermore, the choice of the embedding parameters lead to varying numbers of embedded data that can be obtained from scalar time series. When estimating TE from neural data, we therefore recommend to control the amount of data in one analysis window that is available after embedding and to design experiments accordingly. For example, the presented MEG data set was sampled at 600 Hz, with 137 repetitions of the stimulus on average, which - after embedding - led to 8800 data points per analysis window of 150 ms. In comparison, for simulated data TE was reconstructed correctly for 10000 data points and more. Thus, in our example MEG data set, shorter analysis windows would not have been advisable because of an insufficient amount of data per analysis window for reliable TE estimation. If shorter analysis windows are necessary, they will have to be counterbalanced by a higher number of experimental repetitions.

Thus, the choice of an appropriate analysis window is crucial to guarantee reliable TE estimation, while still resolving the temporal dynamics under investigation. A further data limiting factor is the need for an appropriate embedding of the scalar time series. To embed the time series at a given point $t$, enough history for this sample (embedding dimension times the embedding delay in sample points) has to be recorded. We call this epoch the *embedding window*. The need for an appropriate embedding thus constitutes another constraint for the data necessary for TE estimation. Thus, the choice of an optimal embedding dimension (e.g. through the use of Ragwitz' criterion [71]) is crucial as the use of larger than optimal embedding dimensions wastes available data and may lead to a weaker estimation rate in noisy data [56].

Note, that the embedding window should not be confused with the analysis window. The analysis window strictly describes the data points, for which neighbor statistics enter TE estimation – where neighbor counts may be averaged over an epoch $\Delta_t$ or may come from a single point in time $t$ only. The embedding window however, describes the data points that enter the embedding of a single point in time. Thus, the temporal resolution of TE analysis may still be in single time steps $t$ (i.e. only one time point entering the analysis), even though the embedding window spans several points in time that contain the history for this single point.

### Repeatability of neuronal processes

When applying the ensemble method to estimate TE from neural recordings, we treat experimental repetitions as multiple realizations of the neural processes under investigation. In doing so, we assume stationarity of these processes *over repetitions*. We claim that in most cases this assumption of stationarity is justified for processes concerned with the processing of experimental stimuli and that the assumption also holds for stimulus-independent processes that contribute to neural recordings. We will first present the different contributions to neural recordings and subsequently discuss their individual statistical properties, i.e. their stationarity over repetitions. Note, that the term stationarity refers to the stability of the *probability distribution underlying* the observed realizations of contributions over repetitions and does not require individual realizations to be identical; i.e. stationarity does not preclude a variability in observed realizations, but rather implies some variance in observed realizations, that is reflective of the variance in the underlying probability distribution.

Contributions to neural recordings may either be stimulus-related (*event-related activity*) or stimulus-independent (*spontaneous ongoing activity*). Within the category of event-related activity, contributions can be further distinguished into phase-locked and non phase-locked contributions (the latter is commonly called *induced activity*). Phase-locked activity has a fixed polarity and latency with respect to the stimulus and – on averaging over repetitions – contributes to an event-related potential or field (ERP/F). Phase-locked activity is further distinguished into two types of contributions, that are discussed as mechanisms in the ERP/F-generation (e.g. [97–99]): (1) *additive evoked contributions*, i.e. neural activity that is in addition to ongoing activity and represents the stereotypical response of a neural population to the presented stimulus in each repetition [100–102]; (2) *phase-reset contributions*, i.e. the phase of ongoing activity is reset by the stimulus, such that phase-aligned signals no longer cancel each other out on averaging over repetitions [103–106]. In contrast to these two subtypes of phase-locked activity, induced activity is event-related activity that is not phase-locked to the stimulus, such that latency and polarity vary randomly over repetitions and induced activity averages out over repetitions.

We therefore have to consider four types of contributions to neural recordings: (1) additive evoked contributions, (2) phase-reset contributions, (3) induced contributions and (4) spontaneous ongoing contributions, the last being stimulus-independent. Stationarity can be assumed for all these contributions if no learning effects occur during the experiment. Learning effects may lead to slow drifts, i.e. changing mean and variances, in the recorded signal. Such learning effects may easily be tested for by comparing the first and second half of recorded repetitions with respect to equal variances and means. If variances and means are equal, learning effects can most likely be excluded. Empirically, the stationarity assumption, specifically of phase-locked contributions, can also be verified using a modified independent component analysis recently proposed in [107].

To sum up the statistical properties of different contributions to neural data and their relevance for using an ensemble approach to TE estimation, we conclude that all contributions to neural recordings can be considered stationary over repetitions by default. Non-stationarity over repetitions will only be a problem in paradigms that introduce (slow) drifts or trends in the recorded signal, for example by facilitating learning during the experiment. Testing for drifts may be done by comparing mean and variance in a split-half analysis.

## Relation of the ensemble method to local information dynamics

We will now discuss the relation of the ensemble approach suggested here to the local transfer entropy (LTE) approach of Lizier [4,15]. This may be useful as both approaches at first glance seem to have a similar goal, i.e. assessing information transfer more locally in time. As we will show, the approaches differ in what quantities they localize. From this difference it also follows that they can (and should be) combined when necessary.

In detail, the ensemble approach used here tries to exploit cyclostationarity or repeatability of random processes to obtain multiple PDFs from the different (quasi-) stationary parts of the repeated process cycle, or a PDF for each step in time from replications of a process, respectively. In contrast, local information dynamics localizes information transfer in time (and space) given the PDF of a *stationary* process.

The local information dynamics approach to information transfer computes information transfer for stationary random processes from their joint and marginal PDFs for each process step, thereby fully localizing information transfer in time. The quantity proposed for this purpose is the LTE [15]:

$$LTE(X \rightarrow Y, t, \delta) = log \frac{p(y_t | \mathbf{y}_{t-1}, \mathbf{x}_{t-\delta})}{p(y_t | \mathbf{y}_{t-1})} \qquad (17)$$

LTE relates to TE in the same way Shannon information relates to Shannon entropy – by means of taking an expected value under the common PDF $p(y_t, \mathbf{y}_{t-1}, \mathbf{x}_{t-\delta})$ of the collection of random variables $\{X_t\}, \{Y_t\}$ that form the processes $\mathbf{X}$, $\mathbf{Y}$, which exchange information. Stationarity here guarantees that all the random variables $X_1, X_2, \ldots (Y_1, Y_2, \ldots)$ have a common PDF (as the PDF is not allowed to change over time):

$$TE(X \rightarrow Y, t, \delta) = <LTE(X \rightarrow Y, t, \delta)>_{p(y_t, \mathbf{y}_{t-1}, \mathbf{x}_{t-\delta})} \quad (18)$$

In contrast, the approach presented here does not assume that the random processes $X$, $Y$ are stationary, but that either replications if the process can be obtained, or that the process is cyclostationary. Under these constraints a *local* PDF can be obtained. The events drawn from this PDF may then be analyzed in terms of their average information transfer, i.e. using TE as presented here, or by inspecting them individually, computing LTE for each event. In this sense, the approach suggested here is aimed at extracting the proper local PDFs, while local information dynamics comes into play once these proper PDFs have been obtained. We are certain that both approaches can be fruitfully combined in future studies.

## Relation of the ensemble method to other measures of connectivity for non-stationary data

Linear Granger causality (GC) is – as has been shown recently by [26] – equivalent to TE for variables with a *jointly* Gaussian distribution. Thus, for data that exhibit such a distribution, information transfer may be analyzed more easily within the GC framework. Similar to the ensemble method for TE estimation, extensions to GC estimation have been proposed that deal with non-stationary data by fitting time-variant parameters. For example, Möller and colleagues presented an approach that fitted multivariate autoregressive models (MVAR) with time-dependent parameters to an ensemble of EEG signals [108]. Similar measures, that fit time-dependent parameters in autoregressive models to data ensembles, were used by [109] and [110]. A different approach to dealing with non-stationarity was taken by Leistritz and colleagues [111]. These authors proposed to use self-exciting threshold autoregressive (SETAR) models to model neural time series within a GC framework. SETAR models extend traditional AR models by introducing state-dependent model parameters and allow for the modeling of transient components in the signal.

The presented methods for the estimation of time-variant linear GC may yield a computationally less expensive approach to the estimation of information transfer from an ensemble of data. However, linear GC is equivalent to TE regarding the full recovery of information transfer for data with a *jointly* Gaussian distribution only. For non-Gaussian data, linear GC may fail to capture higher order interactions. As neural data are most likely non-Gaussian, the application of TE may have an advantage for the analysis of information transfer in this type of data. The non-Gaussian nature of neural data can for example be seen, when comparing brain electrical source signals from physical inverse methods to time courses of corresponding ICA components [112]. Here, ICA components and extracted brain signals closely match. Given that ICA components are as non-Gaussian as possible (by definition of the ICA), we can infer that brain signals are very likely non-Gaussian.

We also note that a nonstationary measure of *coupling* between dynamic systems building on repetitions of time series and next-neighbor statistics was suggested by Andrzejak and colleagues [113]. The key difference of their approach to the ensemble method suggested here is that the previous states of the target time series are not taken into account explicitly in their method. Hence, their measure is not (and was not intended to be) a measure of information transfer (see [53] for details why a measure of information transfer needs to include the past state of target time series, and [48] for the difference between measures of (causal) coupling and information transfer). In addition, their methods explicitly tries to determine the *direction* of coupling between systems. This implies that there should be a dominant direction of coupling in order to obtain meaningful results. Transfer entropy, in contrast, easily separates and quantifies both directions of information transfer related to bidirectional coupling, under some mild conditions related to entropy production in each of the two coupled systems [53].

## Relation of the ensemble method to the direct method for the calculation of mutual information of Strong and colleagues

The ensemble method proposed here shares the use of replications (or trials) with the so called 'direct method' of Strong and colleagues [114]. The authors introduced this method to calculate mutual information between a controlled stimulus set and neural responses. Similarities also exist in the sense that the surrogate data method for statistical evaluation used in our ensemble method builds on trial-to-trial variability, as does Strong's method (by looking at intrinsic variability versus variability driven by stimulus changes).

However, the two methods differ conceptually on two accounts: First, the quantity estimated is different – symmetric mutual information in Strong's method compared to inherently asymmetric conditional mutual information in the case of TE. Second, the method of Strong and colleagues requires a direct intervention in the source of information (i.e. the stimuli) to work, whereas TE in general is independent of such interventions. This has far reaching consequences for the interpretation of the two measures: The intervention inherent in Strong's method places it somewhat closer to causal measures such as Ay and Polani's causal information flow [47], whereas intervention-free TE has a clear interpretation as the information transferred in relation to distributed computation [48]. As a consequence, TE maybe easily applied to quantify neural information transfer from one neuron or brain area to another even under *constant* stimulus conditions. In contrast, using Strong's method inside a neural system in this way would require precisely setting of the activity of the source neuron or brain area, something that may often be difficult to do.

## Application of the proposed implementation to other dependency measures

The use of ensemble pooling of observations for the estimation of time-resolved dependency measures has been proposed in a variety of frameworks. For example, Andrzejak and colleagues [113] use ensemble pooling of delay-embedded time series in combination with nearest neighbor statistics as a general approach to the estimation of arbitrary non-linear dependency measures. However, the practical application of ensemble pooling and nearest neighbor statistics together with the necessary generation of a sufficient amount of surrogate data sets (typically > 1000 in neuroscience applications where correction for multiple comparisons is necessary) was always hindered by its high computational cost. Only with the presentation of a GPU algorithm for nearest neighbor searches, we provide an implementation of the ensemble method that allows its practical application. Note that even though we use ensemble pooling and GPU search algorithms to specifically estimate TE, the presented implementation may easily be adapted to other dependency measures that are calculated from (conditional) mutual informations estimated from nearest neighbor statistics.

## Application to MEG source activity in a perceptual closure task

Application of the ensemble-based TE estimation to MEG source activities revealed a time varying pattern of information transfers, as expected in the nonstationary setting of the visual task. While a full discussion of the revealed information transfer pattern is beyond the scope of this study, we point out individual connections transferring information that underline the validity of our results. Notable connections in the first time window transfer information from the early visual cortices (V1) to the orbitofrontal cortex (OFC) – in line with earlier findings by Bar an colleagues [92], that suggest a role of the OFC in early visual scene segmentation and gist perception. Another brain area receiving information from early visual cortex is the caudal inferior temporal gyrus (cITG)[115], an area responsible for the processing of shape-from-shading information, which is thought to be essential for perception of Mooney stimuli as they were used here. Both of these areas, OFC and cITG at later stages of processing exchange information with the fusiform face area, which is essential for the processing of faces [116–118], and thereby expected to receive information from other areas in this task. Indeed, FFA seems to be an essential hub in the task-related network investigated in this study and receives increasing amounts of incoming information transfer as the task progresses in time. This is in line with the fact that the most pronounced task-related differences in FFA activity were found at latencies > 200 ms previously [86].

Our data also clearly show a great variability in information transfer pattern across subjects, which we relate to the limited amount of data per subject, rather than to true variation. Moreover, future investigations will have to show whether more fine grained temporal segmentation of the neural information processing in this task is possible and whether it will provide additional insights.

## Conclusion and further directions

We presented an implementation of the ensemble method for TE presented in [55], that uses a GPU to handle computationally most demanding aspects of the analysis. We chose an implementation that is flexible enough to scale well with different experimental designs as well as with future hardware developments. Our implementation was able to successfully reconstruct information transfer in simulated and neural data in a time-resolved fashion. Nearest neighbor searches using a GPU exhibited substantially reduced execution times. The implementation has been made available as part of the open source MATLAB toolbox TRENTOOL [56] for the use with CUDA-enabled GPU devices.

We conclude that the ensemble method in its presented implementation is a suitable tool for the analysis of non-stationary neural time series, enabling this type of analysis for the first time. It may also be applicable in other fields that are concerned with the analysis of information transfer within complex dynamic systems.

## Author Contributions

Conceived and designed the experiments: MW RV MMZ PW FDP. Performed the experiments: PW MMZ MW. Analyzed the data: PW MMZ MW. Contributed reagents/materials/analysis tools: MMZ PW MW RV FDP. Wrote the paper: PW MMZ RV MW. Conceived and designed the parallel algorithm: MW MMZ. Implemented the algorithm: MMZ PW MW FDP. Designed the software used in analysis: MW RV MMZ PW FDP.